\documentclass[12pt]{article}

\usepackage{amsmath}
\usepackage{amssymb}
\usepackage{graphicx}
\usepackage{appendix} 
\usepackage{cite}
\usepackage{color}
\usepackage{array}
\usepackage{epsfig} 

\newcommand{\R}{{\mathcal R}}
\usepackage[labelfont=bf,labelsep=period,justification=raggedright]{caption}
\makeatletter
\renewcommand{\@biblabel}[1]{\quad#1.}
\makeatother
\date{}

\begin{document}
\baselineskip=20pt 

\begin{flushleft}
\textbf{\Large Early Real-time Estimation of the Basic Reproduction
Number of Emergening Infectious Diseases} \\
Bahman Davoudi$^{1,*}$, Joel C. Miller$^{2}$, Rafael Meza$^{1}$,
Lauren Ancel Meyers$^{3}$, David J. D. Earn$^{4}$, Babak Pourbohloul$^{1,5}$
\\
\textbf{1} Mathematical ModelingServices, British Columbia Centre
for Disease Control, Vancouver, British Columbia, Canada \\
\textbf{2} Department of Epidemiology, Harvard School of Public Health,
Boston, Massachusetts, U.S.A. \\
\textbf{3} Section of Integrative Biology, Institute for Cellular
and Molecular Biology, University of Texas at Austin, Austin, Texas,
U.S.A. \\
\textbf{4} Department of Mathematics \& Statistics and the M. G.
De Groote Institute of Infectious Disease Research, McMaster University,
Hamilton, Ontario, Canada \\
\textbf{5} School of Population \& Public Health, University of British
Columbia, Vancouver, British Columbia, Canada \\
\textbf{{*}}\emph{Corresponding Author} 
\par\end{flushleft}

\section*{Abstract}

When an infectious disease strikes a population, the number of newly
reported cases is often the only available information during the
early stages of the outbreak. An important goal of early outbreak
analysis is to obtain a reliable estimate for the basic reproduction
number, $\R_{0}$. 

Over the past few years infectious disease epidemic processes have
gained attention from the physics community. Much of the work to date,
however, has focused on the analysis of an epidemic process in which
the disease has already spread widely within a population; conversely,
very little attention has been paid, in the physics literature or
elsewhere, to formulating the initial phase of an outbreak. Careful
analysis of this phase is especially important as it could provide
policymakers with insight on how to effectively control an epidemic in
its initial stage.

We present a novel method, based on priciples of network theory, that
enables us to obtain a reliable real-time estimate of the basic reproduction
number at an early stage of an outbreak. Our method takes into account
the possibility that the infectious period has a wide distribution
and that the degree distribution of the underlying contact network
is heterogeneous. We validate our analytical framework with numerical
simulations.

\section*{Introduction}

\label{intro} The basic reproduction number, $\R_{0}$, is a fundamental
characteristic of the spread of an infectious disease. It is generally
defined as the expected number of new infections caused by a typical
individual during the entire period of his/her infection in a totally
susceptible population \cite{Anderson_1991,Anderson_2000,Dietz_1993}.
Because $\R_{0}$ is a simple scalar quantity, and perhaps because
in many circumstances it determines the expected (average) final size
of an outbreak \cite{Kermack_1927,Dietz_1993,Ma_2006,Arino_2007,Brauer_2008},
it has been widely used to gauge the degree of threat that a specific
infectious agent will pose as an outbreak progresses \cite{Hethcote_2000,Lipsitch_2003,Fraser_2009}.
While it is clear that knowing the value of $\R_{0}$ can be very
useful for policymakers in planning a response; it is not as straightforward
to obtain a reliable estimate of $\R_{0}$, especially in early stages
of an outbreak, before large scale uncontrolled transmission has taken
place and before the basic biology and transmission pathways of the
pathogen have been characterized.

Early in an outbreak, the pattern of disease spread is predominantly
influenced by the probabilistic nature of infection transmission.
Consequently, a wide array of outcomes is possible, ranging from the
outbreak fizzling out, even in the absence of an intervention, to
circumstances where the initial stage expands into a large-scale epidemic.
Once a full-blown epidemic develops, several assumptions can be made
that simplify the estimation of $\R_{0}$, as has been discussed in
detail in the literature \cite{Anderson_1991,Dietz_1993,Hethcote_2000,Ma_2006}.

In many cases, it is necessary to assess the impact of various intervention
strategies before a large-scale epidemic occurs. In doing so, stochastic
manifestations of disease transmission as well as the underlying structure
of the contact network should be taken into account. The first aspect
has been widely studied. For example, the Reed-Frost model is a chain-binomial
stochastic model where each infected individual can infect susceptible
individuals and they are all assumed to have the same contact rate
\cite{Abbey_1952,Bailey75,Addy_1991,Ball_1986}. Another example is
the methodology developed by Becker \cite{Becker_1976}, and Ball
and Donnelly \cite{Ball_1995} based on a branching process susceptible-infected-recovered
(SIR) model. Branching processes have received wide attention because
they facilitate the evaluation of the basic reproduction number as
well as the final epidemic size and epidemic probability \cite{Guttorp_1991}.
More recently, the 2003 global outbreak of severe acute respiratory
syndrome (SARS) inspired the development of new methodologies based
on the daily number of new cases and the distribution of the serial
interval between successive infections \cite{Wallinga_2004,Cauchemez_2006_1,Cauchemez_2006_2,White_2008}.
However, none of these methods take into consideration the influence
of the contact network underlying an epidemic process or it is assumed
that the contact network is a classic random graph.

Several new methods to estimate the basic reproduction number,
$\R_{0}$, were proposed during or shortly after the 2009 H1N1 influenza pandemic.
Notably,  Nishiura {\it et al.}\cite{Nishiura_2011} employed an
age-structured model to derive an estimate for ${\cal R}_0$. Katriel
{\it et al.}\cite{Katriel_2009} used a new discrete-time stochastic
epidemic SIR model that explicitly takes into account the disease’s specific generation-time distribution and the
intrinsic demographic stochasticity inherent to the infection
process. Balcan {\it et al.} \cite{Balcan_2009} employed a method that
is based on the distribution of arrival times of the H1N1 in 12 different countries seeded by 
the Mexico epidemic using one million computationally simulated epidemics. 
Nishiura {\it et al.} \cite{Nishiura_2009} also developed a discrete time stochastic model 
that accounts for demographic stochasticity and conditional
measurement and applied to estimate the ${\cal R}_0$ value using the
weekly incidence of H1N1 pandemic influenza in Japan. Although all of
these constitute an important advancement in the literature, none of
them simultaneously addresses analytically the stochasticity due to the underlying
contact network and the transmission process.

\section*{Outline Summary}
In the following we first describe the basic notion of network model. 
We then define \emph{infection hazard or infectivity function}, the \emph{removal
hazard or removal function} and their related quantities namely the transmissibility and the
removal probability density function. Furthermore, we derive a stochastic renewal equation that
allows us to understand the relationship between the rate of newly-infected
individuals at a given time and all preceding infected cases in the initial stage of an outbreak. 
We study the renewal 
equation in the exponential regime and obtain the Wallinga-Lipsitch equation\cite{Wallinga_2007}. 
In this regime, we also obtain 
an equation that express the generation interval distribution in term of the transmissibility 
and the removal probability distribution function. In next part, 
we obtain an equation for the basic
reproduction of removed individual. This equation acts as a constrain allowing us
to find one unknown. We will then draw an algorithm and show how one can evaluate
the basic reproduction number. We finally present our extensive numerical results 
showing the accuracy of the methodology
in predicting the basic reproduction number for different network models and parameters
types and values.

\section*{Network basis}

\label{NB_section} 
This section briefly introduces the idea of contact
network epidemiology and defines the key concepts of infection rate,
removal rate, transmissibility and removal probability density. We map a collection of $N$ individuals
to a network where each vertex represents an individual and each edge
shows a pathway of possible infection transmission between two individuals.
We use $k$ to denote the degree of a given individual (the number
of contacts that s/he has) represented graphically by the number of
edges emanating from a vertex, and we use $p_{k}$ to denote the degree
distribution (the probability that a randomly chosen vertex has degree
$k$ ($k$ contacts)). Several important quantities can be derived
once a network's degree distribution is known. The \emph{moments}
of the degree distribution are $\langle k^{n}\rangle=\sum_{\kappa=0}^{\infty}k^{n}p_{\kappa}$.
For $n=1$, $\langle k\rangle$ is the average number of nearest neighbours
of a randomly chosen individual, which we denote $z_{1}$. The average
number of second nearest neighbours of a randomly selected individual,
$z_{2}$, can be expressed as $\langle k^{2}\rangle-\langle k\rangle$
\cite{Newman_2002}. To estimate $\R_{0}$, we count the number of
edges along which an individual can infect others, once that individual
has become infected. This quantity, usually termed \emph{excess degree,}
represents the number of edges emanating from a vertex (individual),
excluding the edge that was the source of the infection. One can show
that the \emph{average excess degree} ($Z_{x}$) is given by the ratio
$\frac{z_{2}}{z_{1}}$ \cite{Noel_2009}.

\begin{figure}[!ht]
\begin{center}
\includegraphics[scale=0.6]{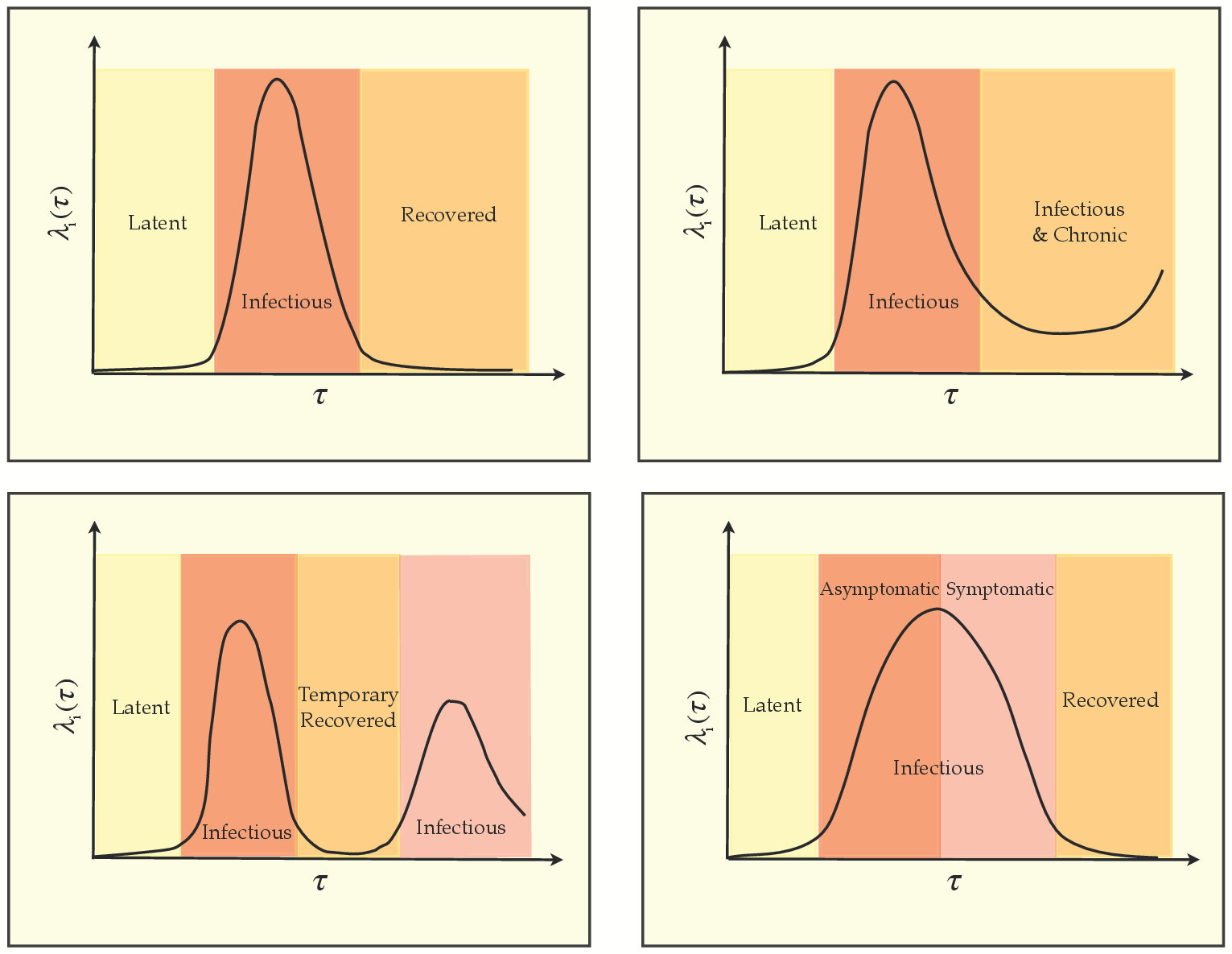}
\end{center}

\caption{Hypothetical \emph{infectivity functions}, $\lambda_{i}(\tau)$. They
show the general infectivity patterns that can occur, varying by complexity
of the disease. Top left panel shows the infectivity function of a
very generic disease, top right panel shows the infectivity function
of an HIV type disease, bottom left displays the infectivity function
of any recurrent disease such as chicken-pox and finally bottom right
panel exhibits the infectivity function an influenza type disease.}

\label{fig_beta(t)} 
\end{figure}

We denote the time at which an individual acquires the infection by
$t_{i}$, and the time since acquiring infection by $\tau=t-t_{i}$
(also known as age of infection). While harbouring the infection,
the individual is first \emph{latent} (infected but not yet infectious)
and then \emph{infectious} (either symptomatically or asymptomatically).
The individual may also \emph{recover}, by which we mean only that
s/he can no longer transmit the infection, not that s/he has necessarily
completely cleared the pathogen. For some diseases, after a temporary
recovery, the person may become infectious again. Knowing that an
individual acquires infection at a given time $t_{i}$, various states
of infectiousness for this individual can be encapsulated within the
infection hazard or \emph{infectivity function}, $\lambda_{i}(\tau)$.
The infectivity function measures the instantaneous risk of disease
transmission across an edge. This implies that for small $\delta\tau$
the conditional probability that infection occurs across an edge between
times $\tau$ and $\tau+\delta\tau$, given that it did not occur
by time $\tau$, can be approximated by $\lambda_{i}(\tau)\delta\tau$.
Typically, $\lambda_{i}(\tau)$ is initially zero during the latent
period, it increases to a certain level and then declines during the
infectious period, before finally vanishing and returning to zero
at the time of permanent recovery. Figure \ref{fig_beta(t)} shows
four hypothetical infectivity functions, the first of which is the
typical case. In practice, the functional form of $\lambda_{i}(\tau)$
should be estimated from the actual transmission profile corresponding
to a specific disease. Note that the only technical restriction on
$\lambda_{i}(\tau)$ is that it must be a non-negative integrable
function.

Given the infectivity function, one can evaluate the probability that
an individual transmits the disease to one of his/her contacts during
a specific time period. Let $T(\tau)$ denote the probability of disease
transmission along one edge for an individual with infection age $\tau$.
Then $T(\tau)$ satisfies \cite{Cox_1984,Newman_2002} 
\begin{equation}
T(\tau)=1-\exp\left(-\int_{0}^{\tau}\lambda_{i}(\tau')d\tau'\right).
\label{eq:Ttr}
\end{equation}

In general, the time-to-removal varies from one individual to another
and there is no \emph{a priori} knowledge of the exact value of this
quantity for each individual. Therefore we must account for its variability
as well. Let $\lambda_{r}(\tau)$ denote the removal hazard or \emph{removal
function}, \textit{i.e.}, the instantaneous rate of removal for an
individual with infection age $\tau$. This implies that for small
$\delta\tau$ the conditional probability that an individual is removed
between times $\tau$ and $\tau+\delta\tau$, given that s/he was
not removed by time $\tau$, can be approximated by $\lambda_{r}(\tau)\delta\tau$.
The removal function indicates how quickly the infectious individuals
are removed from disease dynamics as a function of the duration of
their infection. This can be related to death or various interventions,
such as quarantine, reduction of social activity due to severity of
illness, behaviour change etc.
Let $\Psi(\tau)$ denote the probability that an individual has time-to-removal
$\geq\tau$. Then \cite{Cox_1984}

\begin{equation}
\Psi(\tau)=\exp\left(-\int_{0}^{\tau}\lambda_{r}(\tau')d\tau'\right),\label{eq:psi}
\end{equation}
 subject to the condition $\Psi(\infty)=0$. The removal probability
density function is given by $\psi(\tau)=-\frac{d\Psi(\tau)}{d\tau}$
(or $\Psi(\tau)=\int_{\tau}^{\infty}\psi(\tau')d\tau'$).

Using Equation \eqref{eq:Ttr} and $\psi(\tau)$ (or $\Psi(\tau)$)
one can calculate the expected transmissibility, i.e the probability
of disease transmission - across a given edge:

\begin{eqnarray}
T & = & \int_{0}^{\infty}\psi(\tau)T(\tau)d\tau=\int_{0}^{\infty}\Psi(\tau)\frac{dT(\tau)}{d\tau}d\tau\nonumber \\
 & = & \int_{0}^{\infty}{\lambda_{r}(\tau)e^{-\int_{0}^{\tau}\lambda_{r}(u)du}[1-e^{-\int_{0}^{\tau}\lambda_{i}(u)du}]d\tau}.\label{eq:T}
\end{eqnarray}
 The basic reproduction number, which represents the average number
of infections caused by a typical infected individual in a fully susceptible
population, can be written as the product of the expected excess degree
and the expected transmissibility \cite{Newman_2002} 
\begin{equation}
\R_{0}=Z_{x}T.\label{eq:R0basic}
\end{equation}

\section*{Disease dynamics on networks}

\label{DD_sec} 
In this section we present some examples of the spread
of an infectious agent on a contact network. The pattern of disease
spread on a network can be categorized into three different regimes:
stochastic, exponential and declining. The process of disease spread
is stochastic in nature, given that the disease transmission along
an edge occurs in a probabilistic manner and that the degree of the
next infected individual cannot be determined \emph{a priori}. The
stochastic behaviour is dominant in the initial stage of disease spread
when the number of infectious individuals is comparatively small (stochastic
regime). The effect of stochasticity becomes much less pronounced
when the number of newly infected individuals becomes significant,
and stochastic fluctuations are smoothed out (exponential regime).
Progression of disease spread starts to decline as the cumulative
number of infected cases becomes comparable to the size of the network,
at which point network finite-size effects becomes important (declining
regime)\cite{Noel_2009}.

\begin{figure}[!ht]

\begin{center}
\includegraphics[scale=0.8]{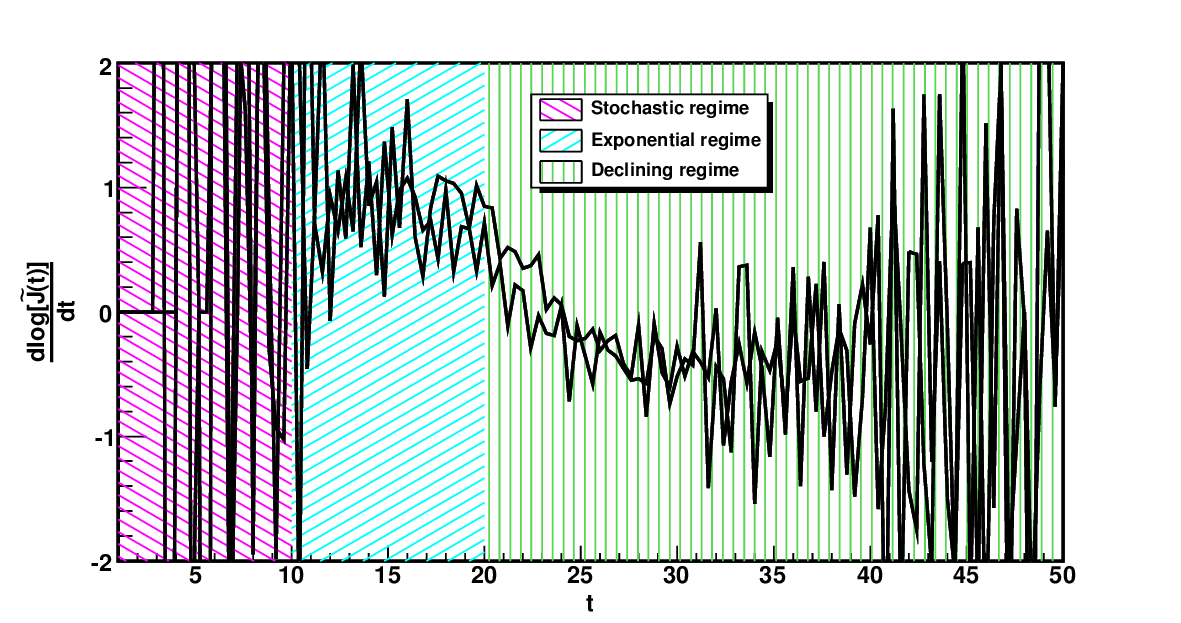}
\end{center}
\caption{Two hypothetical realizations of an epidemic process on a network.
The left, middle and right represent stochastic, exponential and declining
regimes, respectively.}
\label{3regimes} 
\end{figure}

From now on, we use the tilde notation to make the distinction between
the realization of a stochastic process (with tilde) and its mean
field value (without tilde). We define $\tilde{j}(t)$ as the time
series of infection events, which is a sum of delta Dirac functions
located at each infection time. The case count $\tilde{C}(t,\delta t)$
that gives the number of infections between times $t$ and $t+\delta t$
can be expressed as $\tilde{C}(t,\delta t)=\int_{t}^{t+\delta t}\tilde{j}(t')dt'$.
We define $\tilde{J}(t)=\tilde{C}(t,\delta t)/\delta t$ as the incidence
rate of new infections at time $t$, where $\delta t$ is the maximum
time resolution. In the exponential regime, the incidence rate of
new infections grows exponentially and therefore can be expressed
as: 
\begin{equation}
\tilde{J}(t)\simeq J_{0}\exp(\alpha t),\label{eq:J_exp}
\end{equation}
 for some $\alpha>0$. The above-mentioned regimes are represented
in Figure \ref{3regimes}.

\subsection*{Stochastic dynamics of disease}

\label{SD_section}

In this section we outline a general framework to estimate the basic
reproduction number assuming that all information about a specific
realization of the epidemic process up to time $t$ is known. We start
by first deriving a renewal equation for the rate of new infections,
$\tilde{J}(t)$.

\subsubsection*{A renewal equation for $\tilde{J}(t)$}

Let's consider the first person in the population infected with the
disease and assume that her/his infection occurred at time $0$. From
equation \eqref{eq:T}, we can infer that the expected number of infections
that this individual will cause by time $t$ is given by 
\begin{equation}
Z_{x}\int_{0}^{t}\Psi(\tau)\frac{dT(\tau)}{d\tau}d\tau
\end{equation}
 (assuming that his/her excess degree is equal to the average excess
degree). This leads in the limit $t\rightarrow\infty$ to the usual
value of ${\cal R}_{0}=Z_{x}T$. The above expression also implies
that the mean contribution of this individual to the incidence rate
of new infections at his/her infection age $\tau$ is given by 
\begin{equation}
Z_{x}\Psi(\tau)\frac{dT(\tau)}{d\tau}.\label{Jind}
\end{equation}

The above equations can be readily generalized to address the random
process of infection spread on a contact network. In particular, one
can compute the contribution of the individuals infected at time $t-\tau$,
$\tilde{J}(t-\tau)\delta t$, to the number of new infections occurring in
initial stage of an outbreak at time $t$, $\tilde{J}(t)\delta t$, namely
\begin{align}
\tilde{J}(t)\delta t=Z_{x}\int_{0}^{t}\tilde{J}(t-\tau)\delta t\Psi(\tau)\tilde{\Theta}(\tau,t)d\tau.\label{Jt}
\end{align}
where $\tilde{\Theta}(\tau,t)$ denotes the fraction of those edges
where disease is actually transmitted exactly at time $t$. This is a
random function with expectation given by
$\frac{dT(\tau)}{d\tau}$. Please note that $Z_{x}$ and $\Psi(\tau)$ are
a function of $t$ in the most general case. Expression \eqref{Jt} is a generalization of the classical Lotka renewal equation for population growth \cite{Lotka_1939,Kot_2002}, here applied to epidemic dynamics when taking into account the structure of the
underlying contact network and the stochasticity inherent to the transmission
process.

\subsubsection*{Exponential regime and the generation interval distribution}

\label{GI_section} In this section we show the relationship between
our results and the methodology developed by Wallinga and Lipsitch
\cite{Wallinga_2007}. During the exponential regime we can ignore
stochastic fluctuations and replace all quantities with their expected
values. In particular, if we ignore stochastic effects we can re-write
the renewal equation \eqref{Jt} as 
\begin{equation}
J(t)=Z_{x}\int_{0}^{t}J(t-\tau)\Psi(\tau)\frac{dT(\tau)}{d\tau}d\tau\label{eq:WL}
\end{equation}
 where we used $\tilde{\Theta}(\tau,t)\approx\frac{dT(\tau)}{d\tau}$.

Let $\chi(\tau)\equiv Z_{x}\Psi(\tau)\frac{dT(\tau)}{d\tau}$. Equation
\eqref{eq:WL} then takes the simpler form

\begin{equation}
J(t)=\int_{0}^{t}J(t-\tau)\chi(\tau)d\tau,\label{eq:Lotka}
\end{equation}
which is the well-known Lotka renewal equation \cite{Lotka_1939,Kot_2002}.

From the definitions of $\chi(\tau)$, the expected transmissibility
(Eq. \eqref{eq:T}) and the basic reproduction number (Eq. \eqref{eq:R0basic})
we can see that $\int_{0}^{\infty}\chi(\tau)d\tau=\R_{0}$. Substituting
$J(t)\approx\exp(-\alpha t)$ in Eq. \eqref{eq:Lotka} and taking
the limit when $t\rightarrow\infty$ we obtain that 
\begin{equation}
1=\int_{0}^{\infty}\exp(-\alpha\tau)\chi(\tau)d\tau.\label{eq:exp}
\end{equation}
It is worth mentioning that the exact exponential regime can be reached
when $t\rightarrow\infty$ for an infinite-size network and that is
why it is valid to take the limit. This means that there should be
a slight deviation from the exponential behaviour for a {}``finite-size''
system at {}``finite time'' once the outbreak has surpassed the
stochastic regime. Dividing both sides of Eq. \eqref{eq:exp} by $\R_{0}$
we find that \cite{Wallinga_2007} 
\begin{equation}
\frac{1}{\R_{0}}=\int_{0}^{\infty}\exp(-\alpha\tau)\hat{\chi}(\tau)d\tau,\label{R0basiceq}
\end{equation}
where $\hat{\chi}(\tau)=\chi(\tau)/\R_{0}$ is defined as the \emph{generation
interval distribution}. This equation relates the basic reproduction
number to the Laplace transform of the generation interval distribution
in the asymptotic case (infinite-size, infinite-time). Now, in our
formulation the generation interval distribution can be written as

\begin{equation}
\hat{\chi}(\tau)=\frac{\Psi(\tau)\frac{dT(\tau)}{d\tau}}{\int_{0}^{\infty}\Psi(\tau')\frac{dT(\tau')}{d\tau'}d\tau'}.\label{chi}
\end{equation}
 This equation describes how the transmissibility, $T(\tau)$, and
the distribution of the time to removal, $\Psi(\tau)$, determine
together the generation interval distribution.

\subsubsection*{The generation interval distribution for constant parameters}

Equation \eqref{eq:exp} has a simpler form for constant $\lambda_{r}$
and $\lambda_{i}$. This can be obtained by replacing $\psi(\tau)=\lambda_{r}\exp(-\lambda_{r}\tau)$
and $T(\tau)=1-\exp(-\lambda_{i}\tau)$ inside Equation \eqref{eq:exp}
\begin{align}
1 & =Z_{x}\lambda_{i}\int_{0}^{\infty}\exp(-\tau\lambda_{r})\exp(-\alpha\tau)\exp(-\lambda_{i}\tau)d\tau\\
 & =Z_{x}\frac{\lambda_{i}}{\lambda_{i}+\lambda_{r}+\alpha}
\end{align}
 or 
\begin{equation}
\alpha=\left(Z_{x}-1\right)\lambda_{i}-\lambda_{r}.\label{eq:alpha}
\end{equation}
Therefore, we can express the rate of exponential growth of an epidemic
in terms of the mean excess degree ($Z_{x}$), the infectivity ($\lambda_{i}$)
and removal ($\lambda_{r}$) rates. 

Furthermore, we can also explicitly compute the generation interval
distribution (Eq. \eqref{chi}) 
\begin{equation}
\hat{\chi}(\tau)=(\lambda_{i}+\lambda_{r})e^{-(\lambda_{i}+\lambda_{r})\tau}.
\end{equation}
For constant parameters the generation interval is an exponential
random variable with mean $1/(\lambda_{i}+\lambda_{r})$. Using this
fact and Eq.~\eqref{R0basiceq} we obtain the following expression
for $\R_{0}$ 
\begin{equation}
\R_{0}=\frac{\lambda_{i}+\lambda_{r}+\alpha}{\lambda_{i}+\lambda_{r}}=1+\frac{\alpha}{\lambda_{i}+\lambda_{r}}.\label{eq:R0constantrates}
\end{equation}
This equation relates the value of $\R_{0}$ to the rate of growth
during the exponential phase of an epidemic ($\alpha$) the infection
rate ($\lambda_{i}$) and the removal rate ($\lambda_{r}$). Notice
that in contrast to the results obtained from a deterministic SIR
model, where the mean generation interval is equal to the mean duration
of infection \cite{Wallinga_2007}, here we find that the mean generation
interval also depends on the infection rate.

\begin{figure}[!ht]

\begin{center}
\includegraphics[scale=0.8]{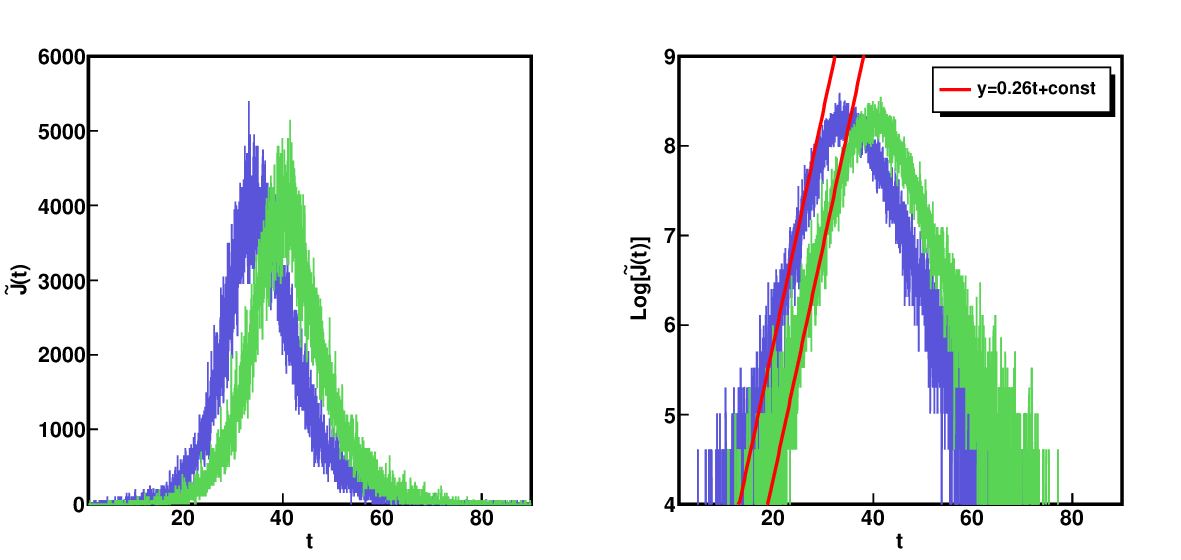} 
\end{center}
\caption{The rate of new infections $\tilde{J}(t)$ (left panel) and its logarithm
(right panel) for a binomial network with $z_{1}=5$, $\lambda_{i}=0.12771$
and $\lambda_{r}=0.25$. Two independent epidemic realizations are
shown in each panel (green and blue). The solid red line shows the
tangent of $Log[\tilde{J}(t)]$ in the exponential regime.}
\label{fig_JlogJ_B} 
\end{figure}

\begin{figure}[!ht]
\begin{center}
\includegraphics[scale=0.8]{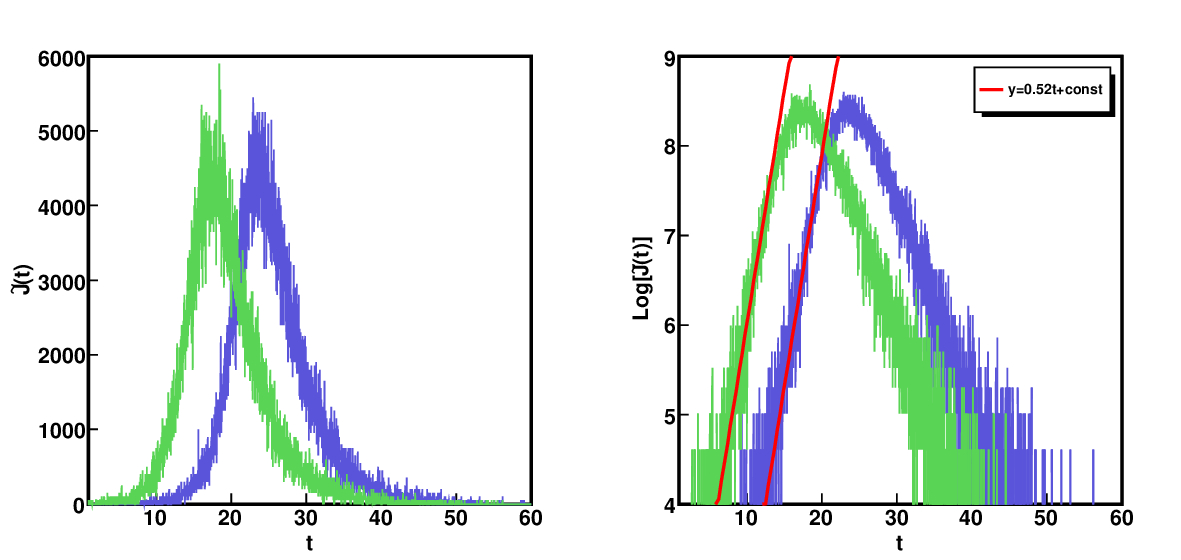} 
\end{center}
\caption{The number of newly-infected cases (left panel) and its logarithm
(right panel) for an exponential network with $\kappa=4$, $\lambda_{i}=0.12771$
and $\lambda_{r}=0.25$. Two independent epidemic realizations are
shown in each panel (green and blue). The solid red line shows the
tangent of $Log[\tilde{J}(t)]$ in the exponential regime.}
\label{fig_JlogJ_E} 
\end{figure}

Figures~\ref{fig_JlogJ_B} and \ref{fig_JlogJ_E} show two examples
of the epidemic behaviour in the three regimes. The algorithm used
to simulate the spread of an infectious agent on a contact network
is described in the appendix. The left panel of Figure \ref{fig_JlogJ_B}
shows two epidemic events unfolding (two different initial index cases)
on a binomial network $p_{k}=\left(\begin{matrix}Nk
\end{matrix}\right)p^{k}(1-p)^{N-k}$ with $N=100,000$ nodes, $z_{1}=5$ (or $p=z_{1}/(N-1)$), $Z_{x}=5$
and $\{\lambda_{i}=0.12771,\;\lambda_{r}=0.25\}$. Using Equation
\eqref{eq:T} the expected transmissibility can be calculated as $T=0.338$.
The stochastic, exponential and declining regimes are approximately
between $0\leq t<20$, $20\leq t<40$ and $40\geq t$ time units,
respectively. Also, in the right panel of Figure \ref{fig_JlogJ_B},
we show the variation of $Log[\tilde{J}(t)]$ with $\delta t=0.02$
over time. The two solid lines $y_{1},y_{2}=\alpha t+const$ with
$\alpha=0.26084$ are the tangent of $Log[\tilde{J}(t)]$ during the
exponential regime (Eq.~ \eqref{eq:alpha}). 
This shows the consistency between the simulated epidemic
curve and its expected (exponential growth) behaviour. Figure \ref{fig_JlogJ_E}
shows the same results for an exponential network with $p_{k}=\left(1-e^{-1/\kappa}\right)e^{-k/\kappa}$,
$\kappa=4$, $z1=3.52$, and $Z_{x}=7.0416$. In this example, the
stochastic, exponential and declining regimes are approximately between
$0\leq t<10$, $10\leq t<20$ and $20\leq t$ time intervals, respectively.
The lines $y_{1},y_{2}=\alpha t+const$ with $\alpha=0.52158$ are
again the tangent of $Log[\tilde{J}(t)]$ during the exponential regime.
Notice that although here we know the {}``true'' value of $\alpha$,
in practice it can be estimated from real-life time series data if
the outbreak progresses beyond the stochastic regime. As can be seen
in Figures \ref{fig_JlogJ_B} and \ref{fig_JlogJ_E} the cases count
$\tilde{J}(t)$, generally resembles a typical epidemic curve if the
pattern of disease spread has a chance to grow substantially. 

\subsubsection*{The number of infected and removed individuals at time $t$}

\begin{figure}[!ht]
 \begin{center}
\includegraphics[scale=0.6]{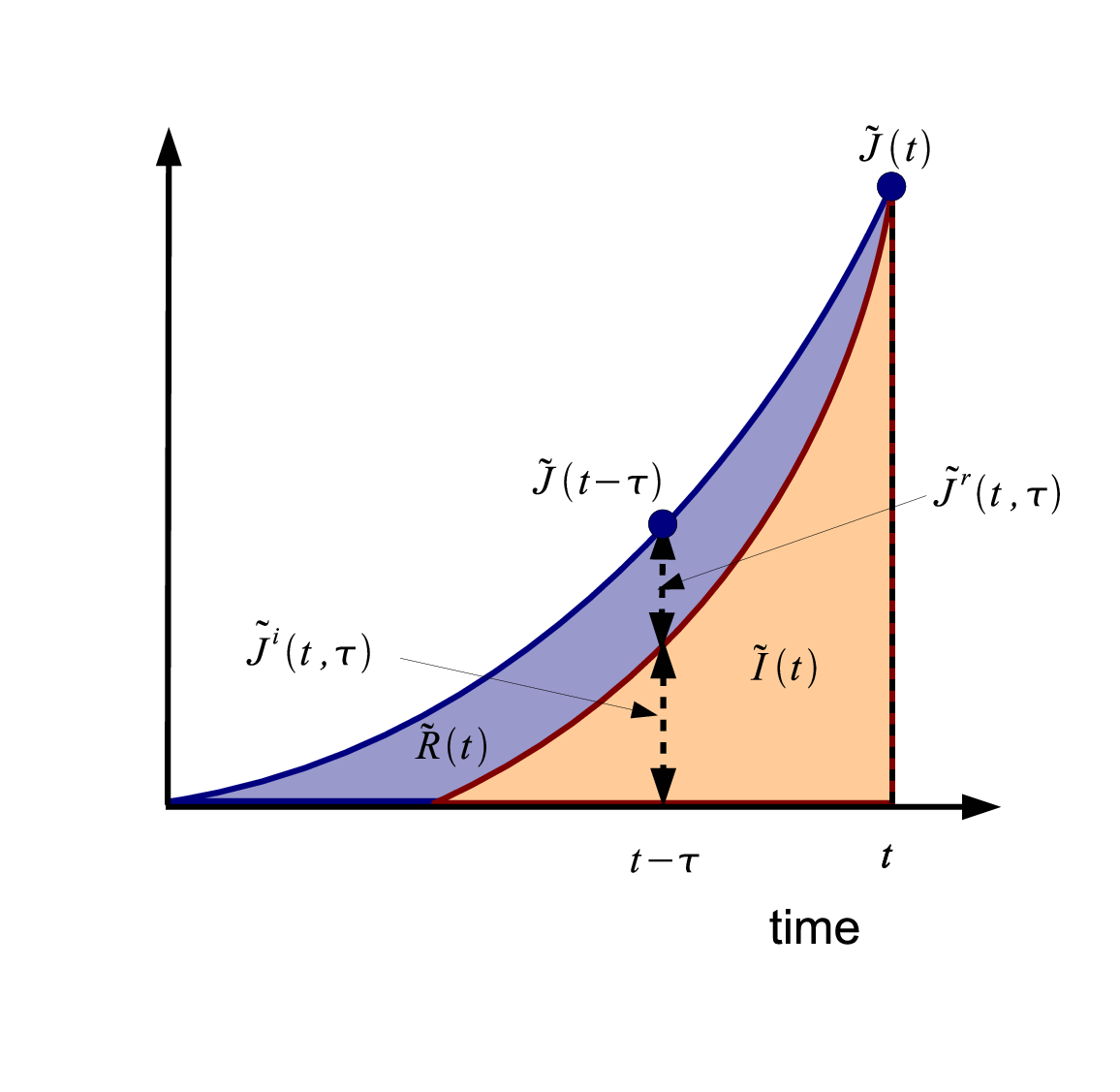}
\end{center}
\caption{This figure illustrates schematically the dependency of the rate of
new infections, $\tilde{J}(t)$ (blue curve), on its past values.
Only a fraction of the cases infected at time $t-\tau$, $\tilde{J}(t-\tau)\delta t$,
contributes to the infections at time $t$, $\tilde{J}^{i}(\tau,t)\delta t=\tilde{J}(t-\tau)\delta t\Psi(\tau)$
(red curve); the rest, $\tilde{J}^{r}(\tau,t)\delta t=\tilde{J}(t-\tau)\delta t[1-\Psi(\tau)]$,
have already been removed.}
\label{JJ} 
\end{figure}

Using the quantities introduced in the previous sections, we now derive
other expressions that will be helpful in estimating the basic reproduction
number, ${\cal R}_{0}$. As an outbreak progresses, at any given time
there is a population of infectious individuals, $\tilde{I}(t)$,
and a population of removed individuals, $\tilde{R}(t)$. The number
of \emph{affected} individuals - the total number of infected individuals
at time $t$ and those who are recovered/removed by time $t$ - is
given by

\begin{align}
\tilde{I}(t)+\tilde{R}(t)=N-\tilde{S}(t)=\int_{0}^{t}\tilde{J}(t-\tau)d\tau,
\end{align}
 where $\tilde{S}(t)$ denotes the number of susceptible individuals
at time $t$. As illustrated in Figure~\ref{JJ}, the total number
of infected cases can be written as

\begin{align}
\tilde{I}(t)=\int_{0}^{t}\tilde{J}^{i}(\tau,t)d\tau=\int_{0}^{t}\tilde{J}(t-\tau)\tilde{\Psi}(\tau)d\tau\label{It}
\end{align}
which in turn, implies that 

\begin{align}
\tilde{R}(t)=\int_{0}^{t}\tilde{J}^{r}(\tau,t)d\tau=\int_{0}^{t}\tilde{J}(t-\tau)\left[1-\Psi(\tau)\right]d\tau
\label{Rt}
\end{align}
where $J^{i}(\tau,t)=\tilde{J}(t-\tau)\Psi(\tau)$ and
$J^{r}(\tau,t)=\tilde{J}(t-\tau)[1-\Psi(\tau)]$.

\begin{figure}[!ht]
\begin{center}
\includegraphics[scale=0.8]{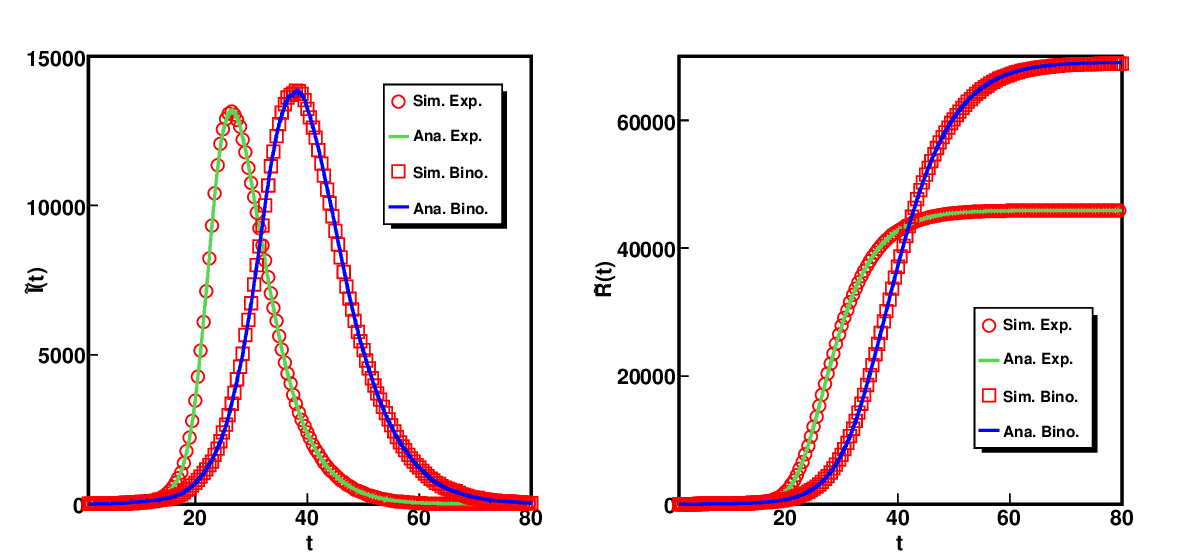} 
\end{center}
\caption{Number of infectious (left panel) and removed individuals (right panel)
as a function of time for the binomial ($z_{1}=5$, $\lambda_{i}=0.12771$
and $\lambda_{r}=0.25$) and exponential ($\kappa=4$) networks. The
solid curves come from the evaluation of Equations \eqref{It}
and \eqref{Rt}), and the symbols from direct counting of
the infectious and removed individuals at any given time for a specific
realization of the process.}
\label{fig_IR_BE} 
\end{figure}

Figure~\ref{fig_IR_BE} shows the number of infectious (left panel)
and removed (right panel) individuals for a disease that spreads either
on a binomial or an exponential network. In both panels the curves
comprising of red symbols correspond to the computer simulation of
an epidemic on the corresponding network; during the simulation the
new case counts are recorded to create a synthetic {}``time series''
for $\tilde{J}(t)$. The solid curves correspond to Equations \eqref{It}
or \eqref{Rt} (for $\tilde{I}(t)$ or $\tilde{R}(t)$) for
the corresponding network. These figures show a perfect agreement
for both networks between the analytical formulas and the case counts
from the simulation.

\begin{figure}[!ht]
\begin{center}
\includegraphics[scale=0.6]{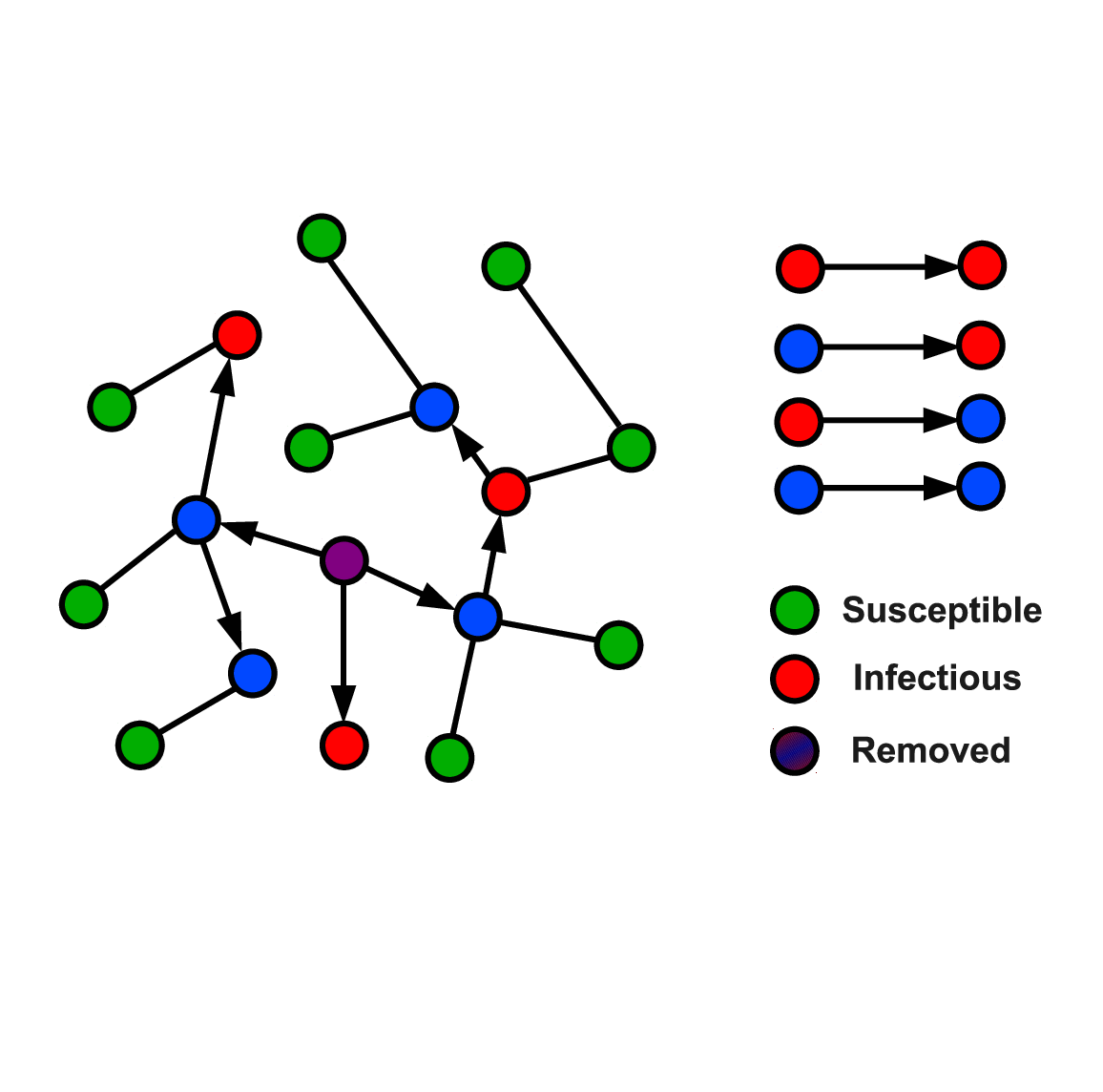}
\end{center}
\caption{A snapshot of the initial phase of an outbreak starting from the central
vertex shown by the purple circle (already removed). The blue/purple,
red and green circles represent removed, infectious and susceptible
individuals, respectively. The links are presented by lines, those
that carried the disease are shown by the arrows. Four different types
of pairs (pair of infected individuals) can be easily recognized in
this figure and they are shown separately on the right hand side of
the figure.}
\label{fig_gg} 
\end{figure}

\begin{figure}[!ht]
\begin{center}
\includegraphics[scale=0.5]{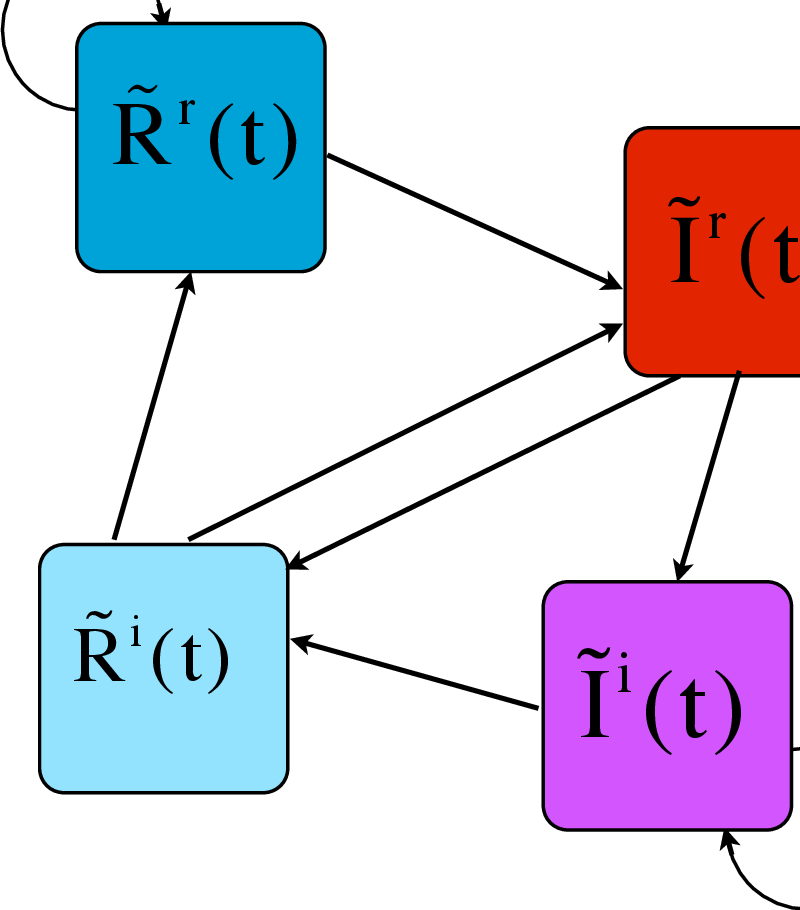} 
\end{center}
\caption{This figure illustrates various options of causality between infected
and recovered cases. The main symbol ($\tilde{I}$ or $\tilde{R}$)
denotes the current status of an individual, and the superscript ($i$
or $r$) corresponds to the status of the person who passed the infection
on to this individual. $\tilde{I}^{r}(t)+\tilde{R}^{r}(t)$ constitutes
the number of infections caused by already removed individuals.}
\label{fig_RRII} 
\end{figure}

In Figure~\ref{fig_gg}, we show the state of the population during
the initial phase of an outbreak starting from the central vertex
shown by the purple circle (already removed). The blue/purple, red
and green circles represent removed, infectious and susceptible individuals,
respectively. The links are presented by lines, those that carried
the disease are shown by the arrows. Four different types of pairs
(pairs of individuals who have been infected) can be easily recognized
in this figure and they are presented on the right side of the figure.
These pairs can be collected in four different groups Figure \ref{fig_RRII} and
can be shown by $\tilde{I}^{i}(t)$, $\tilde{I}^{r}(t)$, $\tilde{R}^{i}(t)$
and $\tilde{R}^{r}(t)$, where $R$ and $I$ refer to the number of
removed and infectious individuals respectively, with two different
predecessors, shown with the superscripts $\{r$,$i\}$ (removed and
infectious). Not all groups are necessarily connected to one another.
For instance, a recovered/removed predecessor, denoted by superscript
$r$ in $\tilde{R}^{r}(t)$, may only come from a group whose members
were recovered while their predecessors were still infectious ($\tilde{R}^{i}$),
or from a group in which the person and his/her predecessor are removed
(denoted by the self-loop from $\tilde{R}^{r}$ to itself). Similarly,
the recovered predecessor ($r$) of an infected person ($\tilde{I}^{r}(t)$),
may only come from a group of recovered individuals whose respective
predecessors were infectious ($\tilde{R}^{i}(t)$) or from a group
whose respective predecessors were already recovered ($\tilde{R}^{r}(t)$).

\subsection*{The transmissibility of removed individuals}
\label{TRI}

Our methodology to estimate ${\cal R}_{0}$ is based on a detailed
analysis of the characteristics of removed individuals. This is because
the history of removed individuals contains all the information about
the mechanisms of disease transmission and recovery process. In particular,
the period of infection of these individuals can help us estimate
the distribution of removal times. Furthermore, since removed individuals
have already had the oppertuinity to transmit the disease, the fraction
of their contacts that they actually infected contains a lot of information
about the transmissibility of disease. In an ideal world, a full characterization
of the infection history of each removed individual would be enough
to estimate ${\cal R}_{0}$. However, in reality, it is extremely
difficult to know which infected individuals have already been removed
and what fraction of their potential infections actually occurred.
Therefore, we instead derive theoretical expressions that can help
us estimate some of these quantities.

First we write an expression for the total number of secondary contacts
of those individuals already removed by time $t$. Using ideas similar
to those above, the total excess degree of \emph{removed} individuals
can be written as 
\begin{align}
\tilde{{\cal Z}}_{x}^{r}(t)=Z_{x}\int_{0}^{t}\tilde{J}(t-\tau)\left[1-\Psi(\tau)\right]d\tau\label{eq:z_tilde}
\end{align}

Here, $\tilde{{\cal Z}}_{x}^{r}(t)$ represents the total number of
edges of already removed individuals that could have transmitted the
infection by time $t$. However, only a fraction of these links actually
transmitted the disease successfully. This latter fraction is given
by $\tilde{I}^{r}(t)+\tilde{R}^{r}(t)$, where $\tilde{I}^{r}(t)$
and $\tilde{R}^{r}(t)$ represent the number of infectious and removed
individuals at time $t$, respectively whose predecessor is already
removed. The ratio of these two quantities represents the fraction
of potential transmissions that actually occurred, which we shall
refer to as the expected \emph{transmissibility of removed individuals}
or $\tilde{T}^{r}(t)$. 
\begin{align}
\tilde{T}^{r}(t)\equiv\frac{\tilde{I}^{r}(t)+\tilde{R}^{r}(t)}{\tilde{{\cal Z}}_{x}^{r}(t)}.
\label{TT}
\end{align}

Estimates for the expected values for these quantities
can, in principle, be calculated based on the rate of new infections,
$J(t)$, using arguments similar to those above. These expressions
are derived in the next section. Equations \eqref{TT} and \eqref{eq:exp} form a set of  equations that allows us to find two unknowns, lets say, the amplitude and variance of infectivity profile that is the subject of other study. Here we use equation \eqref{TT} to solely estimate the basic reproduction number, $\R_{0}$.
 
\subsection*{Expression for the transmissibility of removed individuals,
$\tilde{T}^{r}(t)$}

As discussed in the previous section, our methodology is based on
a careful analysis of the characteristics of the removed individuals.
In particular, the expected \emph{transmissibility of removed individuals},
$\tilde{T}^{r}(t)$, will play a crucial role in the estimation of
$\R_{0}$. We now derive an alternative expression for $\tilde{T}^{r}(t)$
and other expressions related to equation \eqref{TT}.

The expected transmissibility of removed individuals, $\tilde{T}^{r}(t)$,
can also be obtained as an extension of Equation \eqref{eq:T} by
replacing the removal distribution, $\psi(\tau)$, with the conditional
distribution of removal time, given that it occurred before time $t$,
defined as $\tilde{\psi^{r}}(\tau,t)$. The quantity $\tilde{\psi^{r}}(\tau,t)\delta\tau$
is proportional to the number of individuals already removed by time
$t$ that were removed after exactly $\tau$ units of time, \textit{i.e.},
$\tilde{\psi^{r}}(\tau,t)\delta\tau\propto\psi(\tau)\int_{0}^{t-\tau}\tilde{J}(\tau')d\tau'\delta\tau$.
This probability function, after incorporating the proper normalization,
can be written as

\begin{equation}
\tilde{\psi}^{r}(\tau,t)=\frac{\psi(\tau)\int_{0}^{t-\tau}\tilde{J}(\tau')d\tau'}{\int_{0}^{t}\psi(\tau')\int_{0}^{t-\tau''}\tilde{J}(\tau'')d\tau'd\tau''}.\label{eq:psi_r}
\end{equation}
 The expected transmissibility of removed individuals can then be
calculated as (see Equation~\eqref{eq:T})

\begin{align}
\tilde{T}^{r}(t)=\int_{0}^{t}\tilde{\psi}^{r}(\tau,t)\tilde{T}(\tau,t)d\tau.\label{Tr}
\end{align}
 where $\tilde{T}(\tau,t)$ is an extension of $T(\tau)$ that takes
into account the stochastic effects in the disease transmission process
(represented by the explicit dependence of this quantity on $t$).

\subsection*{The basic reproduction number of removed individuals and an equation
for $\tilde{\R}_{0}$}

The total excess degree of removed individuals (given
in equation \eqref{eq:z_tilde}) takes the simpler form: 
\begin{equation}
\tilde{{\cal Z}}_{x}^{r}(t)=Z_{x}\tilde{R}(t).\label{eq:z_simple}
\end{equation}
 Combining the last equation with equation \eqref{TT} one obtains
\begin{align}
\tilde{T}^{r}(t)Z_{x}=\frac{\tilde{I}^{r}(t)+\tilde{R}^{r}(t)}{\tilde{R}(t)}.\label{R0r}
\end{align}
 We define the right hand side of the previous equation as the \emph{reproduction
number of the removed individuals} $\tilde{\R_{0}^{r}}(t)$, i.e.,
\begin{equation}
\tilde{\R_{0}^{r}}(t)=\tilde{T}^{r}(t)Z_{x}.\label{eq:r_removed}
\end{equation}
 Using Equations \eqref{R0r} and \eqref{eq:r_removed}, we can write
a time-dependent estimator of the basic reproduction number, $\tilde{\R}_{0}=Z_{x}T$:
\begin{align}
\tilde{\R}_{0}(t)=\frac{\tilde{I}^{r}(t)+\tilde{R}^{r}(t)}{\tilde{R}(t)}\frac{T}{\tilde{T}^{r}(t)}.
\label{R0}
\end{align}

\begin{figure}[!ht]
\begin{center}
\includegraphics[scale=0.35]{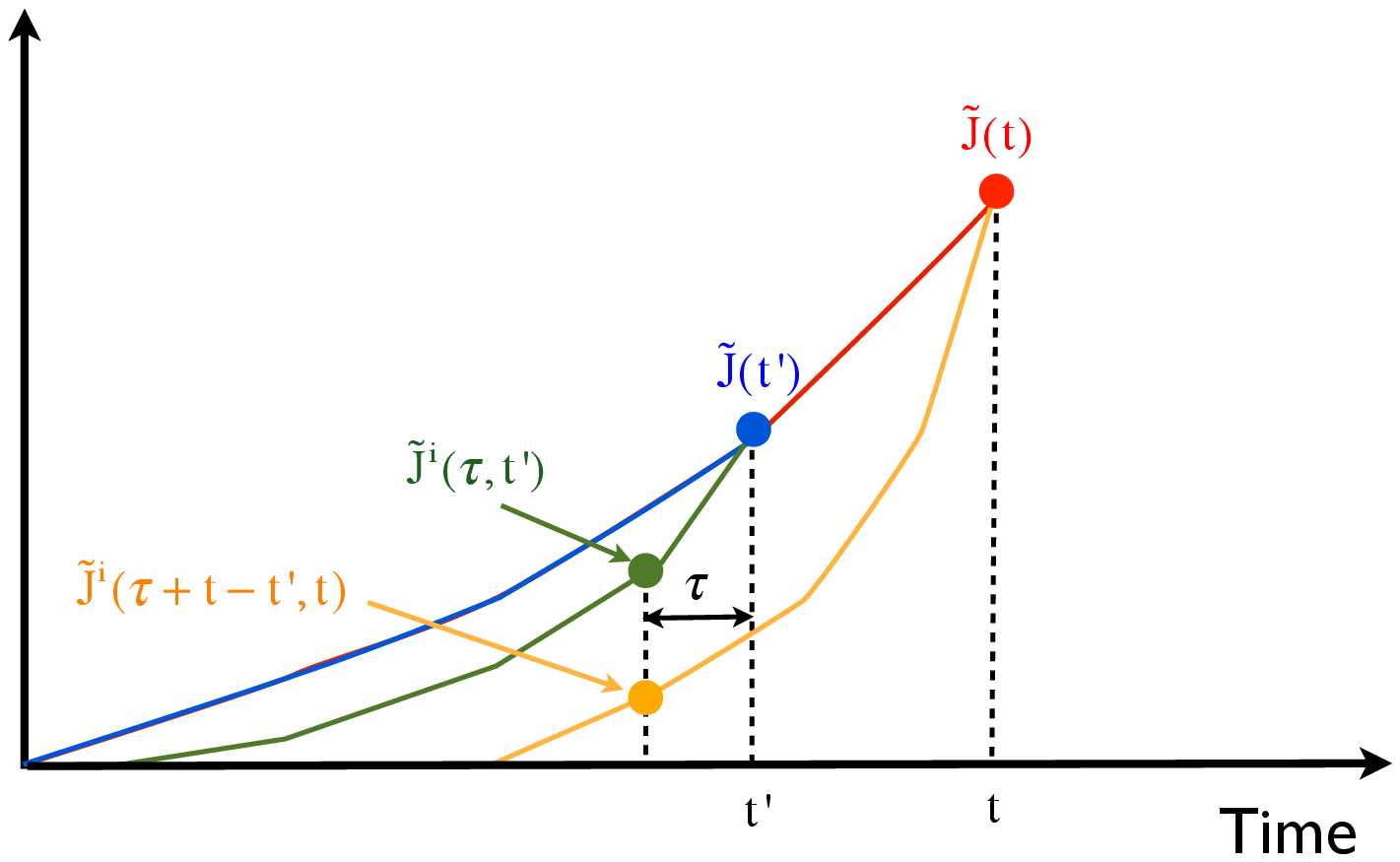} 
\end{center}
\caption{Dependency of the rate of new infections at time $t'$, $\tilde{J}(t')$,
on the rate of new infections at time $t'-\tau$, $\tilde{J}^{i}(\tau,t')$.
The blue and red curves show the rates of new infections by time $t'$
and $t$, respectively. The green and yellow curves show the fraction
of those that remained infectious for at least $\tau$ and $\tau+t-t'$
units of time, respectively.}
\label{fig_Ji} 
\end{figure}

On the right hand side of the last expression, the expected value
of $\tilde{I}^{r}(t)+\tilde{R}^{r}(t)$ can be calculated as 
\begin{align}
\tilde{I}^{r}(t)+\tilde{R}^{r}(t)=\int_{0}^{t}\int_{0}^{t'}\tilde{J}(t')\tilde{\eta}(\tau,t',t)\tilde{\zeta}(\tau,t')d\tau dt'\label{IRr}
\end{align}
where $\tilde{\eta}(\tau,t',t)$ is the fraction of infected individuals
who are removed by time $t$ and who may have infected others at time
$t'\leq t$. $\tilde{\eta}(\tau,t',t)$ can be written as 
\begin{align}
\tilde{\eta}(\tau,t',t)=\frac{\tilde{J^{i}}(\tau,t')-\tilde{J^{i}}(\tau+t-t',t)}{\tilde{J^{i}}(\tau,t')}=\frac{\Psi(\tau)-\Psi(\tau+t-t')}{\Psi(\tau)},
\end{align}
where $\tilde{J^{i}}(\tau,t')=\tilde{J}(t'-\tau)\Psi(\tau)$. Figure~\ref{fig_Ji}
illustrates how the new infection rate at time $t'$ depends on the
infection rate at time $t'-\tau$.

$\tilde{\zeta}(\tau,t')$ is the probability function that an infection
at time $t'$ was caused by any of these individuals. $\tilde{\zeta}(\tau,t')$
can be written as 
\begin{align}
\tilde{\zeta}(\tau,t')=\frac{\tilde{J^{i}}(\tau,t')\tilde{\Theta}(\tau,t')}{\int_{0}^{t}\tilde{J^{i}}(\tau',t')\tilde{\Theta}(\tau',t')d\tau'}.
\end{align}
 
Substituting the expressions of $\tilde{\eta}(\tau,t',t)$ and $\tilde{\zeta}(\tau,t)$
in Equation~\eqref{IRr} we obtain 
\begin{align}
\tilde{I}^{r}(t)+\tilde{R}^{r}(t)=\int_{0}^{t}\tilde{J}(t')\left[1-\frac{\int_{0}^{t}\tilde{J^{i}}(\tau+t-t',t)\tilde{\Theta}(\tau,t)d\tau}{\int_{0}^{t}\tilde{J}^{i}(\tau,t')\tilde{\Theta}(\tau,t)d\tau}\right]dt',
\end{align}
 Notice that the right-hand side depends only on the rate of new infections,
$\tilde{J}(t)$, and disease transmissibility (represented by $\tilde{\Theta}(\tau,t)$).
It is also important to notice that outcome of Equation~(\ref{R0}) is invariant under 
$J(t)\longrightarrow {\rm const}*J^{rep}(t)$ where $J^{rep}(t)$ is rate of new reported cases.
Finally, we notice that for a disease with the constant removal function,
$\lambda_{r}$, $\tilde{J^{i}}(\tau+t-t',t)=\tilde{J}(t'-\tau)\Psi(\tau)\Psi(t-t')$.
Therefore, $\tilde{I}^{r}(t)+\tilde{R}^{r}(t)=\tilde{R}(t)$. This
means that the first fraction on the right hand side in Equation~\eqref{R0r}
equals unity, and thus $\tilde{\R_{0}^{r}}(t)=\tilde{T}^{r}(t)Z_{x}=1$.
The expression for $\tilde{\R}_{0}(t)$ then takes the simpler form:
\begin{align}
\tilde{\R}_{0}(t)=\frac{T}{\tilde{T}^{r}(t)}.\label{R0e}
\end{align}

\subsection*{An algorithm for the estimation of the basic reproduction number}

Using the expressions derived in the previous section, we can compute
real time estimates of the basic reproduction number, ${\cal R}_{0}$,
assuming some knowledge of the underlying contact network and certaincharacteristics
of the disease. For example, if we assume that we know the rate of
new infections up to time $t$ ($J(s),$ $s\le t$), the average excess
degree of the underlying contact network, $Z_{x}$, and the removal
time density, $\psi(\tau)$, we can calculate an estimator of ${\cal R}_{0}$
as follows.
\begin{enumerate}
\item {Evaluate the conditional distribution of removal time given that
it occurred before time $t$, $\tilde{\psi}^{r}(\tau,t)$, using Equation
\eqref{eq:psi_r} and $J(t)$.} 
\item {Calculate $\tilde{T}(\tau,t)$ by equating the left- and right-hand
side of Equation \eqref{R0r}. We should use equation \eqref{Tr}
to evaluate the left-hand side of \eqref{R0r}. It is worth mentioning
that since we use only one equation, we can estimate only one parameter.
This means that we must assume a functional form for $\tilde{T}(\tau,t)$
that depends on at most one parameter value. For example, we could
assume that $\tilde{T}(\tau,t)={\cal T}(\tau)\tilde{{\cal A}}(t)$,
where ${\cal T}(\tau)$ denotes the dependence of the transmissibility
on the age of infection and $\tilde{{\cal A}}(t)$ denotes an amplitude
effect that captures the stochasticity of the transmissibility as
a function of time. Assuming that ${\cal T}(\tau)$ is given, then
$\tilde{T}(\tau,t)$ depends only on the multiplicative parameter
$\tilde{{\cal A}}(t)$.} 
\item {Calculate an estimator of the expected transmissibility, $\hat{T}(t)$,
using $\tilde{T}(\tau,t)$, $\psi(\tau)$ and Equation \eqref{eq:T}.
Notice that the dependence of $\hat{T}(t)$ on $t$ denotes that we
are using only information up to time $t$. } 
\item {The estimated reproduction number at time $t$ is given by ${\cal \hat{R}}_{0}(t)=Z_{x}\hat{T}(t)$.} 
\end{enumerate}
The above algorithm can be modified depending on the information available
for the estimation. For example, if there is enough empirical evidence
to determine the distribution of the duration of infectiousness as
well as the recovery rate of individuals in advance, then the methodology
can be used to shed light on the structure of the underlying contact
network by estimating $Z_{x}$. Examples of this and other applications
of the methodology are given below. But first, a demonstration of
the theoretical aspects of our analytical framework. For more details please see \cite{Pourbohloul_2009}

\section*{Numerical results}

To test the framework presented so far, we performed epidemic spread
simulations on the two networks introduced earlier (binomial and exponential)
and in each case collected the {}``time series'' of case counts
resulting from the simulations.

\subsection*{Constant infectivity and removal functions}

\begin{figure}[!ht]
\begin{center}
\includegraphics[scale=0.8]{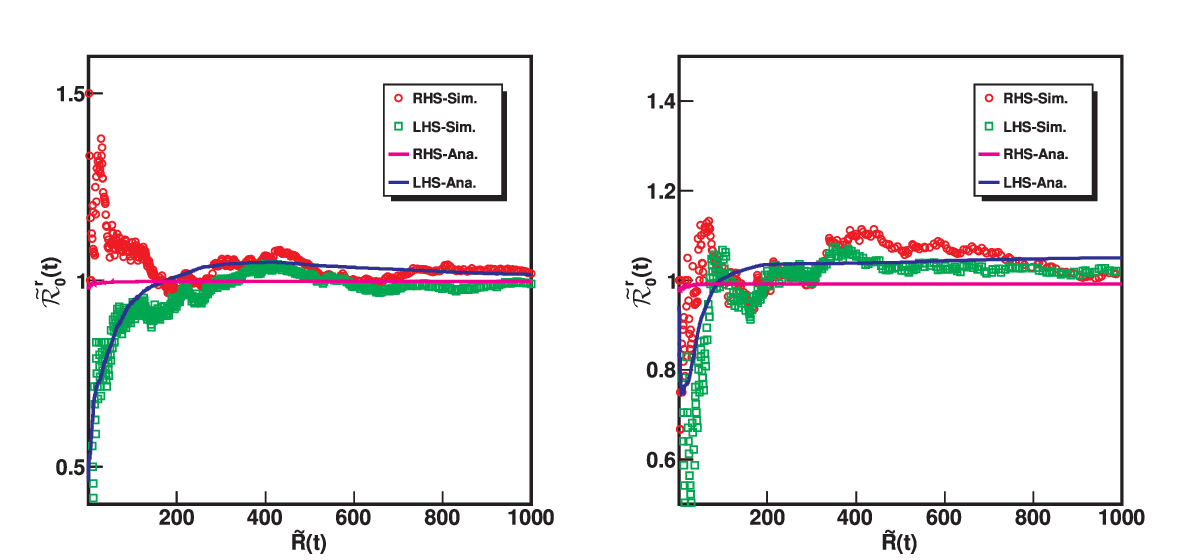} 
\end{center}
\caption{The estimated basic reproduction number of removed individuals for
the binomial (left panel; $z_{1}=5$, $\lambda_{i}=0.12771$ and $\lambda_{r}=0.25$)
and exponential (right panel; $\kappa=4$) networks in terms of the
number of removed individuals. RHS and LHS represent the right and
left hand side of Eq. \eqref{R0r}, respectively. The solid curve
represents the {}``analytical'' calculation of Eq. \eqref{R0r}
and the symbols show the {}``exact'' values of RHS and LHS for this
specific realization of the epidemic process.}
\label{fig_R0r} 
\end{figure}

In Figure~\ref{fig_R0r} we present the basic reproduction number
of the removed individuals, $\R_{0}^{r}(t)$, as a function of the
number of removed individuals, $R(t)$. The symbols show the results
from direct counting during the simulation, whereas the lines show
the results obtained from analytically evaluating (after setting $\tilde{T}(\tau,t)=T(\tau)$
and $\tilde{\Theta}(\tau,t)=\frac{dT(\tau)}{d\tau}$ in the left-
and right-hand sides of Equation \eqref{R0r}) each of the terms in
Equation~\eqref{R0r}. The green/blue and red/pink colors correspond
to the left and right hand side of Equation~\eqref{R0r}, respectively.
The small asymptotic deviation of our estimate for
the exponential network comes from finite size effects \cite{Noel_2009}
(for more details, see figure \ref{Flo:z2z1} in the appendix).

\begin{figure}[!ht]
\begin{center}
\includegraphics[scale=0.7]{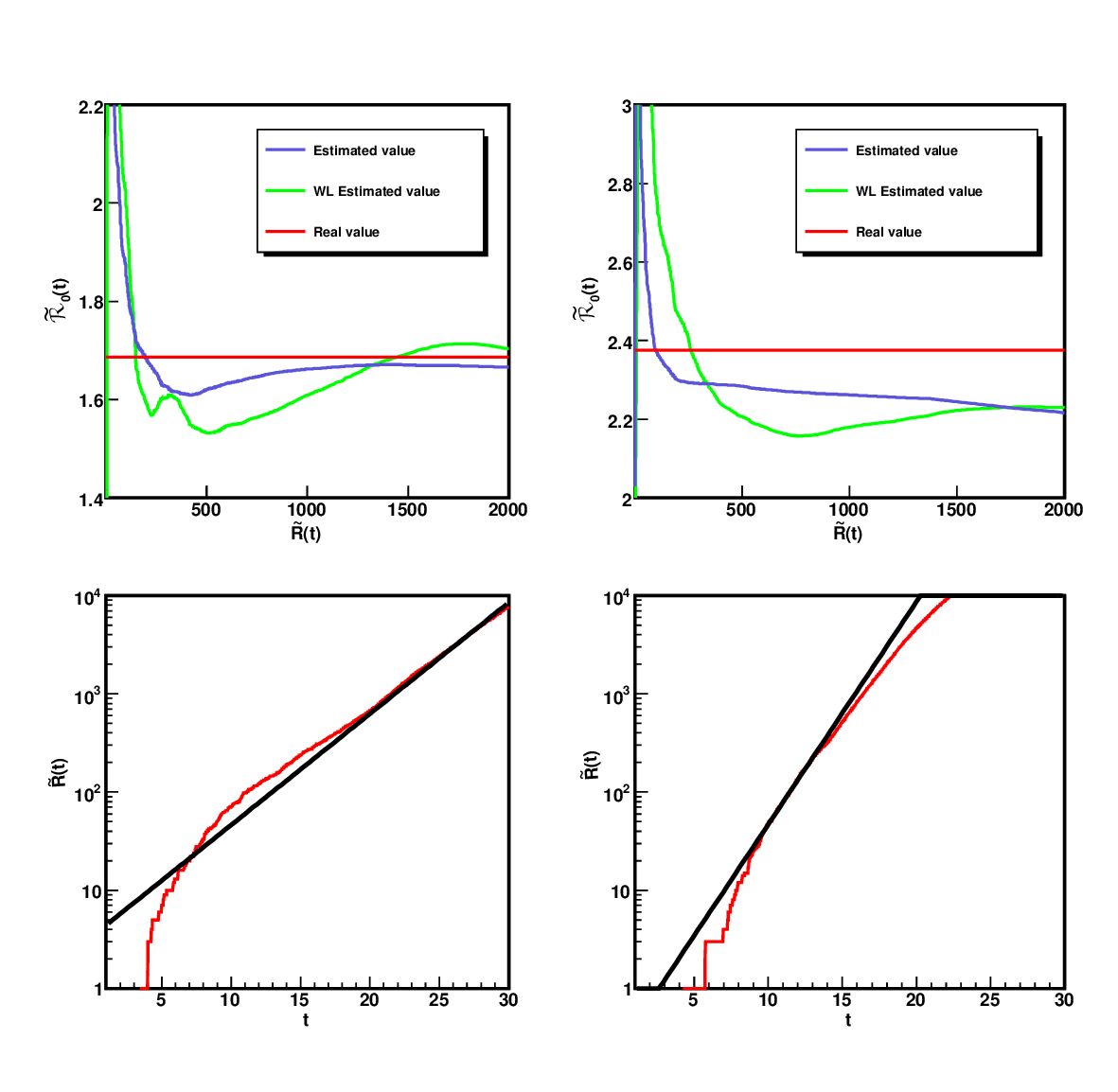} 
\end{center}
\caption{\textbf{Top panels.} Estimated basic reproduction number for the binomial
(left panel; $z_{1}=5$, $\lambda_{i}=0.12771$ and $\lambda_{r}=0.25$)
and exponential (right panel; $\kappa=4$) networks in terms of the
number of removed individuals. The red line corresponds to the real
value of the basic reproduction number. The blue line shows the estimated
$\R_{0}$ from our methodology. For comparison, the green line shows
the $\R_{0}$ estimates obtained from Eq. \eqref{eq:R0constantrates}
and estimation of the growth rate during the exponential phase ($\alpha$).
\textbf{Bottom panels.} The number of removed individuals by time
$t$ (logarithmic scale). The red line shows the number of removed
individuals from a realization of the epidemic process. For comparison,
the black line shows the theoretical exponential phase of the epidemic
process.}
\label{fig_R0} 
\end{figure}

Figure~\ref{fig_R0} shows our estimated basic reproduction number
(blue line) for the binomial (left panel) and exponential (right panel)
networks as a function of the number of removed individuals available
by time $t$. The{}``true'' values (red line) are $\R_{0}=1.6864$
and $2.3749$, respectively. For comparison, the figure also shows
the $\R_{0}$ estimates obtained from equation Eq. \eqref{eq:R0constantrates},
which is equivalent to the Wallinga-Lipsitch (WL) methodology. To
compute the WL estimates we require knowledge of the epidemic exponential
phase's growth rate ($\alpha$). For each simulation, we estimated
$\alpha$ from the logarithm of the cumulative incidence using simple
linear regression and a window of four units of time of data for each time
point. The figure shows that in both cases our estimator converges
and becomes stable quicker than the WL estimator. This is because
our methodology does not make an explicit assumption of exponential
epidemic growth and is therefore able to incorporate and appropriately
weight the information from the stochastic phase of the epidemic.

\begin{figure}[!ht]
\begin{center}
\includegraphics[scale=0.8]{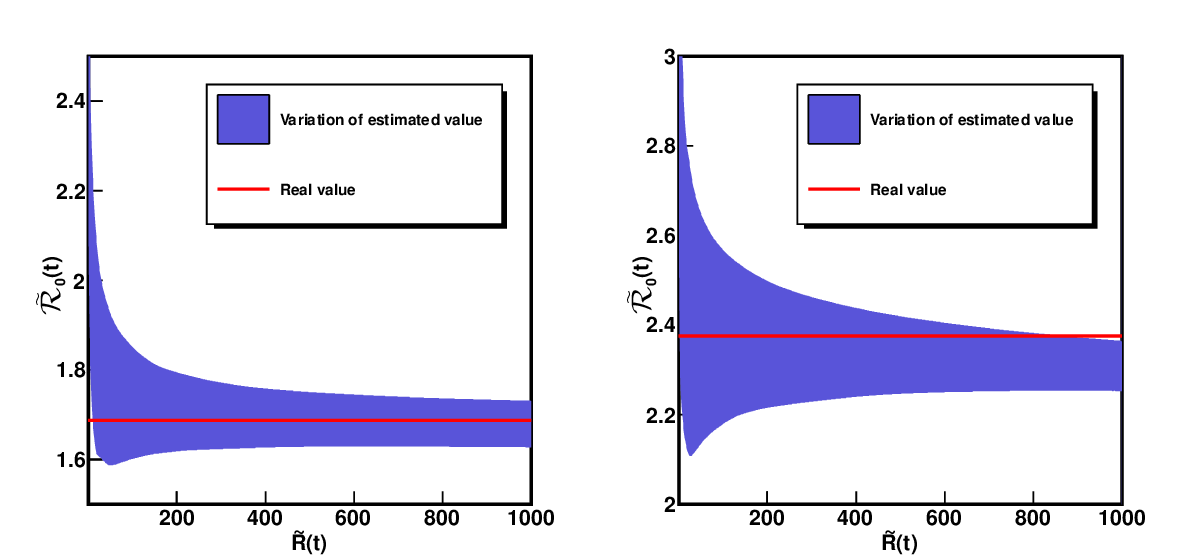} 
\end{center}
\caption{The estimated basic reproduction number for the binomial (left panel;
$z_{1}=5$, $\lambda_{i}=0.12771$ and $\lambda_{r}=0.25$) and exponential
(right panel; $\kappa=4$) networks in terms of the number of removed
individuals. The red line corresponds to the real value of the basic
reproduction number. The blue area shows the variation of the estimated
value for a hundred different realizations.}
\label{fig_R0variability} 
\end{figure}

In order to assess the variability of our estimator, we simulated
one hundred different realizations of the epidemic process and then
estimated the value of $\R_{0}$ for each of them. Figure~\ref{fig_R0variability}
shows the mean estimated value plus/minus one standard deviation (averaging
across realizations) for each network. Notice that the variability
for the exponential network is bigger than for the binomial network.
We attribute this to the fact that the exponential degree distribution
has a larger variance. In addition, the $\R_{0}$ estimate for the
exponential network also appears to have a negative bias. As mentioned
above, we attribute this phenomena to finite size effects, which are
stronger for this network in comparison to the binomial.

It is worth noting that $\tilde{T}^{r}(t)$ is a function of $\lambda_{i}(\tau)$,
$\lambda_{r}(\tau)$ and $J(t)$ (see Equations \eqref{eq:T}, \eqref{eq:psi_r}
and \eqref{Tr}). Therefore, when $\lambda_{r}$ is constant, the
condition $\tilde{T}^{r}(t)Z_{x}=1$ allows one to evaluate, for a
given time series, one of the three quantities $\lambda_{i}$, $\lambda_{r}$
or $Z_{x}$, if the other two quantities are assumed to be known (this
statement holds even if $\lambda_{r}$ is a function of time, in which
case the more complex Equation \eqref{R0r} should be used). Examples
of the estimation of $\lambda_{i}$ or $Z_{x}$ when the other quantities
are assumed to be known are shown in the following.

As mentioned earlier $\tilde{T}^{r}(t)$ is a function of
$\lambda_{i}(\tau)$, $\lambda_{r}(\tau)$ and $J(t)$ (see Equations
\eqref{eq:T}, \eqref{eq:psi_r} and \eqref{Tr}). Therefore, when
$\lambda_{r}$ is constant, the condition $\tilde{T}^{r}(t)Z_{x}=1$
allows one to evaluate, for a given time series, one of the three
quantities $\lambda_{i}$, $\lambda_{r}$ or $Z_{x}$, if the other
two quantities are assumed to be known (this statement holds even
if $\lambda_{r}$ is a function of time, in which case the more complex
Equation \eqref{R0r} should be used). For instance, we simulated
again the epidemic on the binomial and exponential networks presented
earlier, but this time with constant values for $\lambda_{r}$ and
$\lambda_{i}$. Using these values and the derived/calculated $\tilde{J}(t)$,
we calculated the value of $Z_{x}$, which is presented in Figure
(\ref{Flo:z2z1}) for the binomial (left panel) and exponential (right
panel) networks. The green lines show our estimates using the above
condition, while the red lines represent the true values. The blue
symbols are the result of the direct count from simulation. It is
interesting to note that the excess degree of infected individuals
in the simulation very quickly tends to the average value for the
binomial network. This explains in part the excellent agreement of
our estimate and the true basic reproduction number for the binomial
network shown in figure~\ref{fig_R0}. However for the exponential
network, the excess degree of infected individuals in the simulation
has a higher variability and does not agree as well with the corresponding
average excess degree. This is mainly due to finite size effects,
which in this case cause an excess degree of infected individuals
lower than the average value. And this produces in turn biases in
the estimation of $\R_{0}$, like those shown in figure~\ref{fig_R0},
and emphasizes the importance of heterogeneity effects for a network
such as the exponential. Finally we should mention that all the discrepancies
discussed above can be removed, or at least reduced, if one incorporates
the true average excess degree from the simulation (blue dots) in
Equation \eqref{Tr}, instead of a theoretical average excess degree.
This could be done if detailed data on the transmission chain and
the contacts of infected individuals during an epidemic were available.

\begin{figure}[!ht]
\begin{center}
\includegraphics[scale=0.8]{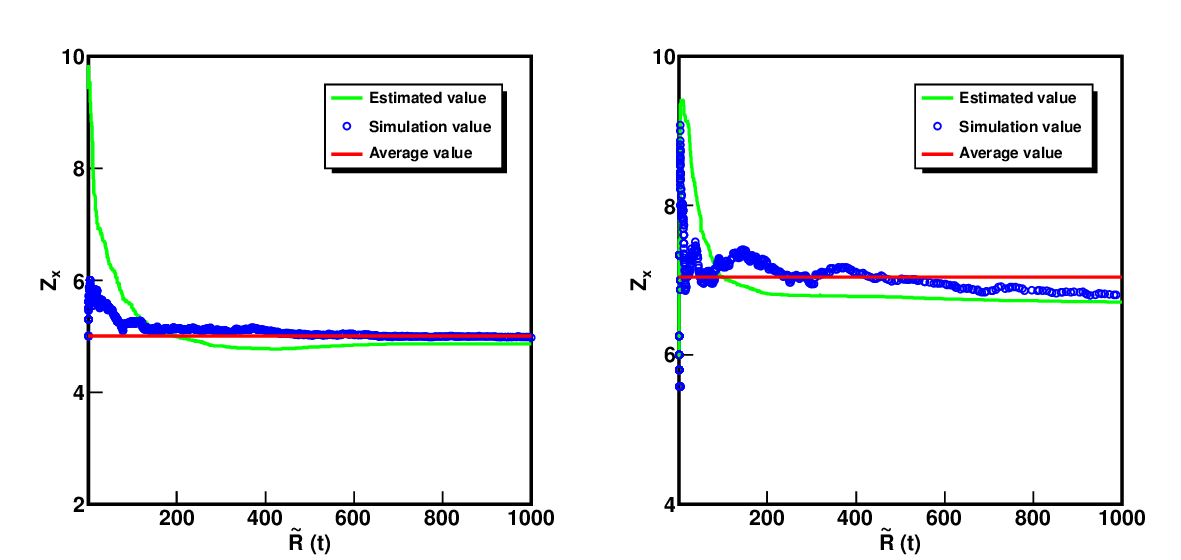}
\end{center}
\caption{Estimates for the value of $Z_{x}$ for the binomial (left) and exponential
(right) networks, when $\lambda_{i}$ and $\lambda_{r}$ are known.
The red line and green curves correspond to the average and the estimated
excess degree.}
\label{Flo:z2z1} 
\end{figure}

\subsection*{Time-dependent infectivity and removal functions}

\begin{figure}[!ht]
\begin{center}
\includegraphics[scale=0.8]{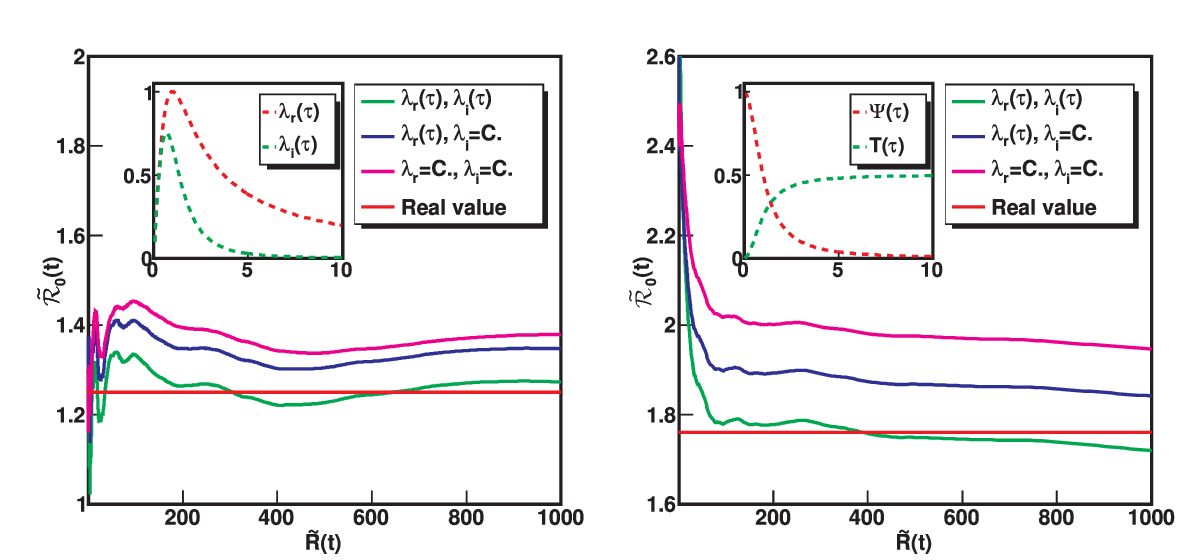}
\end{center}
\caption{The estimated basic reproduction number for the binomial (left panel;
$z_{1}=5$, $\lambda_{i}=\tau/(1+\tau^{2})(1+0.5\tau^{2})$ and $\lambda_{r}=2\tau/(1+\tau^{2})$)
and exponential (right panel; $\kappa=4$) networks in terms of the
number of removed individuals. The red line and green curves correspond
to the real and the estimated value. The blue line shows the estimates
if we incorrectly assume that $\lambda_{i}$ is constant. The pink
line shows the corresponding estimates if we incorrectly assume that
both $\lambda_{i}$ and $\lambda_{r}$ are constant}

\label{fig_R0td} 
\end{figure}

As another example, we use time-dependent infectivity and removal
functions to simulate the epidemic process. Specifically, we assume
that $\lambda_{i}(\tau)=\tau/(1+\tau^{2})(1+0.5\tau^{2})$ and $\lambda_{r}(\tau)=2\tau/(1+\tau^{2})$.
These choices lead to the following expressions for $T(\tau)$ and
$\Psi(\tau)$, $T(\tau)=0.5\tau^{2}/(1+\tau^{2})$ and $\Psi(\tau)=1/(1+\tau^{2})$
(or $\psi(\tau)=2\tau/(1+\tau^{2})^{2}$). Figure \ref{fig_R0td}
shows the estimated reproduction number for the binomial (left panel;
$z_{1}=5$) and exponential (right panel; $\kappa=4$) networks in
terms of the number of removed individuals. The real values in this
case are $1.25$ for the binomial and $1.76$ for the exponential
network. The true values are shown in red and the estimated values
in green. As discussed above, the estimation of the basic reproduction
number is performed as follows (assuming that $\lambda_{r}(\tau)$
and $Z_{x}$ are known). 1) For each $t$, we use Equation~\eqref{R0r}
with $\tilde{T}(\tau,t)=\frac{\tilde{A}(t)\tau}{1+\tau^{2}}$ and
$\tilde{\Theta}(\tau,t)=\frac{2\tilde{A}(t)\tau}{(1+\tau^{2})^{2}}$
and then find the $\tilde{A}(t)$ so that the left- and right-hand
sides of the equation are equal. 2) We then use the calculated $\tilde{A}(t)$
to obtain the expected transmissibility using Equation \eqref{eq:T}.
Finally we calculate the basic reproduction number, ${\cal R}_{0}=Z_{x}T$.
This approach can be used for a specific disease to find the amplitude
of the infectivity function, $\tilde{A}(t)$, assuming we know the
dependence of $\tilde{\Theta}(\tau,t)$ (or $\tilde{T}(\tau,t)$)
on the age of infection, $\tau$. Figure \ref{fig_R0td} also shows
the estimated values of $\R_{0}$ that we get if we assume (erroneously)
instead that either $\lambda_{i}$ is constant (blue curve) or that
both $\lambda_{i}$ and $\lambda_{r}$ are constant (pink curve).
The results show that although the methodology is sensitive to misspecifications
in the functional forms of $\lambda_{i}$ and $\lambda_{i}$, the
estimated $\R_{0}$ values are still relatively close to the true
value.

\subsection*{Sensitivity Analysis}

\begin{figure}[!ht]
\begin{center}
\includegraphics[scale=0.8]{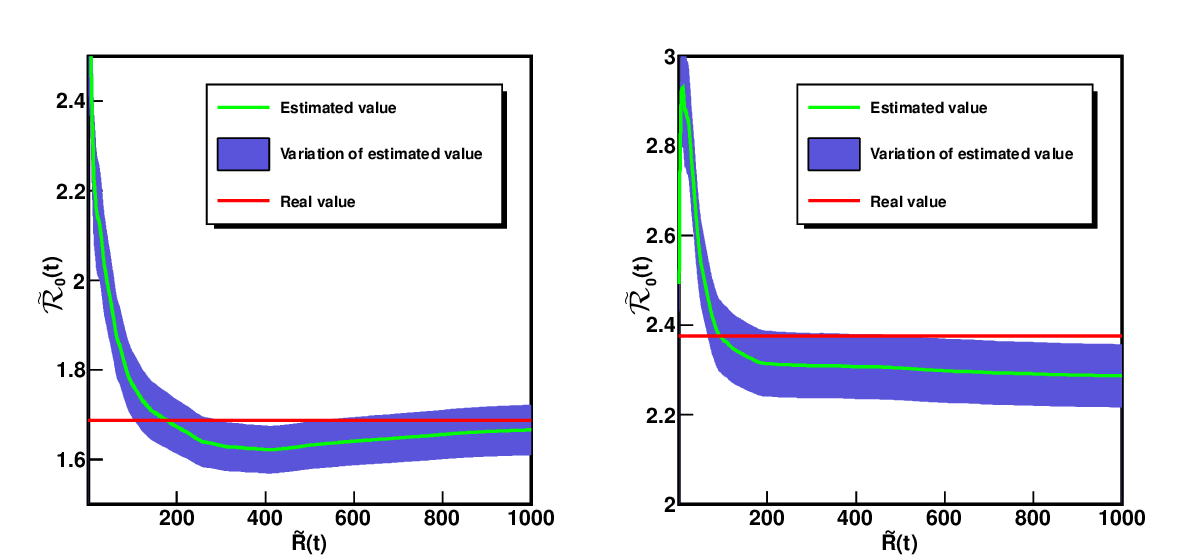} 
\end{center}
\caption{The estimated basic reproduction number for the binomial (left panel;
$z_{1}=5$, $\lambda_{i}=0.12771$ and $\lambda_{r}=0.25$) and exponential
(right panel; $\kappa=4$) networks in terms of the number of removed
individuals. The green curve and red line correspond to the estimated
(with the correct excess degree) and real value of the basic reproduction
number. The blue area shows the variation of the estimated value with
respect to change of the excess degree.}
\label{fig_R0z} 
\end{figure}

Figure \ref{fig_R0z} shows the sensitivity of the estimated reproduction
number to misspecifications of the excess degree. To test this, we
vary the assumed excess degree between $\{4-6\}$ for the binomial
network (true value equal to $5$) and between $\{6.04-8.04\}$ for
the exponential network (true value equal to $7.04$). The results
show that although we assumed a misspecification in the excess degree
of up to $20\%$ for the binomial and up to $14.2\%$ for the exponential
network, the estimates had an error of at most $3\%$ and $3.1\%$,
respectively. As described earlier in the description of the algorithm,
we can use Equation \eqref{R0r} to estimate one model parameter,
in this case $\lambda_{i}$. And then use its value to evaluate the
expected transmissibility and then the expected reproduction number
(the first identity in Equation \eqref{R0}). This shows the usefulness
of Equation \eqref{R0r}, which acts as a strong constraint that allows
us to estimate the basic reproduction number with good precision regardless
of misspecification in other input parameters.

\subsection*{Discussion}

Using concepts from network theory and stochastic processes, we presented
a method that provides a reliable estimate of the basic reproduction
number, $\R_{0}$. Our method takes into account the stochasticity
in disease spread and does not make an explicit assumption about exponential
epidemic growth and therefore is able to provide estimates of $\R_{0}$
at an early stage of an outbreak (i.e., before the exponential regime).
We provided the details of calculations and compared our results at
each step against simulations. Case notification data (time series)
is the main input to this analytical framework. As an outbreak begins
to unfold, the pattern of spread depends substantially on the structure
of the underlying contact network. In fact, this dependency manifests
itself in the formation of the time series of newly infected cases.
The proposed methodology highlights the interplay between the heterogeneity
in contacts (network structure), estimates of the basic reproduction
number and infection transmissibility. Depending on the circumstances,
this methodology can be used to infer other useful quantities as well.
For infectious pathogens that cause repeated outbreaks, there is enough
empirical evidence to establish the distribution of the duration of
infectiousness as well as the recovery rate of individuals. In this
case, in addition to the basic reproduction number, the proposed methodology
can shed light on the structure of the underlying contact network
by estimating the mean excess degree $Z_{x}$. This is an important
piece of information, because in many circumstances it is not possible
to capture and build a detailed contact network among individuals
based on some network generative rules. The importance of this quantity
becomes more apparent when an emerging infectious disease strikes
a population. In this circumstance, there is much less information
on the characteristics of the disease such as the duration of infectiousness
and recovery rate, which in turn determine the transmissibility of
disease. Knowledge of disease transmissibility during the early stage
of an epidemic can play a crucial role, as effective and cost-effective
public health intervention strategies hinge on the degree of contagiousness
of a disease. On the other hand, before the spread of disease becomes
rampant, the structure of the contact network within a population
remains more or less stable. Therefore, the estimated value of $Z_{x}$
obtained during epidemic lulls, from the time series corresponding
to common infections, can be used to estimate the transmissibility
of an emerging infectious disease at the early stage of an outbreak.
We demonstrated this concept with two examples. Our estimate for the
basic reproduction number converges quickly, thus enabling epidemiologists
and policymakers to identify the optimal control strategies, in real
time and even before or at the beginning of the exponential growth
of an epidemic.

\section*{Acknowledgments}

BP would like to acknowledge the support of the Canadian Institutes
of Health Research (grant nos. MOP-81273, PPR-79231 and Team Leader
grant (CanPan II)), the Michael Smith Foundation for Health Research
(Senior Scholar Funds) and the British Columbia Ministry of Health
(Pandemic Preparedness Modeling Project). BD, JM and RM were supported
by these grants. DE was supported by CIHR, NSERC and the J.S. Mc Donnell
Foundation.


\newpage{}

\appendix

\section*{Appendix}

\subsection*{Simulation Algorithm}

To perform Monte Carlo simulations of an epidemic propagation on a
contact network, one first requires explicit knowledge of the network
structure. In this article we use the method described in \cite{Newman_2002,Noel_2009}
to produce a contact network, given a specific degree distribution.
Briefly, (i) sample a random degree sequence ${k_{j}}$ of length
$N$ from the degree distribution ${p_{k}}$, (ii) make sure that
$\sum_{j}k_{j}$ is an even number since a link is composed of two
stubs by reducing the degree of a random individual by one if necessary,
(iii) for each $j$, produce a node with $k_{j}$ stubs, (iv) randomly
choose a pair of unconnected stubs and connect them together; repeat
until all unconnected stubs are exhausted, (v) test for the presence
of self-loops and repeated links. Remove the faulty stubs by randomly
choosing a pair of connected stubs and rewire them by switching stubs.
Repeat until no self-loops and/or repeated links are found.

To simulate the spread of disease on a contact network in continuous
time we follow a Tau-Leaping approach \cite{Gillespie_2001,Gillespie_2003,Higham_2008},
which we describe below. The processes of disease transmission along
one link and the removal of infectious individuals are controlled
by $\lambda_{i}(\tau)$ and $\lambda_{r}(\tau)$, respectively. We
divide time into intervals of length $\Delta t$ and ensure that $\lambda_{i}(\tau)\Delta t$
and $\lambda_{r}(\tau)\Delta t$ are small enough, such that the expected
epidemic curve does not vary much by reducing $\Delta t$ even further.
At every $\Delta t$ step, each infectious individuals recovers with
probability $\lambda_{r}(\tau_{j})\Delta t$, where $\tau_{j}$ is
the age of infection of individual $j$. If an infectious individual
does not recover, then s/he infects independently each of his/her
susceptible contacts with probability $\lambda_{i}(\tau_{j})\Delta t$.

\begin{table}[ht]
\caption{Notation}
\begin{centering}%
\begin{tabular}{|p{4cm}|p{1.5cm}|p{8cm}|}
\hline 
&  & \tabularnewline
Quantity  & Symbol  & Description\tabularnewline
&  & \tabularnewline
&  & \tabularnewline
\hline 
Degree distribution  & $p_{k}$  & Probability that a randomly chosen vertex has degree $k$. \tabularnewline
\hline 
Average degree  & $z_{1}$  & Average degree of vertices in a network calculated by $z_{1}=\left\langle k\right\rangle $.\tabularnewline
\hline 
Excess degree  & $Z_{x}$  & Average degree of a vertix chosen by sampling an edge (calculated
by $Z_{x}=\left\langle k^{2}-k\right\rangle $ $\left\langle k\right\rangle $).\tabularnewline
\hline 
Infection hazard or Infectivity function & $\lambda_{i}(\tau)$  & Instantaneous rate of infection. $\lambda_{i}(\tau)\delta\tau$ gives
the probability of disease transmission across an edge between infection
age $\tau$ and $\tau+\delta\tau$, given it occurred after age $\tau$.\tabularnewline
\hline 
Removal hazard or removal function & $\lambda_{r}(\tau)$  & Instantaneous removal rate. $\lambda_{r}(\tau)\delta\tau$ gives the
removal probability of an infectious individual between its infection
age $\tau$ and $\tau+\delta\tau$, given it occurred after age $\tau$. \tabularnewline
\hline 
Transmissibility & $T(\tau)$  & Probability of disease transmission by infection age $\tau$. It is
calculated by $T(\tau)=1-\exp\left(-\int_{0}^{\tau}\lambda_{i}(\tau')d\tau'\right)$.\tabularnewline
\hline 
Removal distribution  & $1-\Psi(\tau)$  & $\Psi(\tau)$ gives the probability of not being removed by age of
infection $\tau$. $\Psi(\tau)=\exp\left(-\int_{0}^{\tau}\lambda_{r}(\tau')d\tau'\right)$.\tabularnewline
\hline 
Removal probability density & $\psi(\tau)$  & $\psi(\tau)=-d\Psi(\tau)/d\tau$.\tabularnewline
\hline 
Expected transmissibility  & $T$  & Probability of disease transmission along one edge, $T=\int_{0}^{\infty}\psi(\tau)T(\tau)d\tau$.\tabularnewline
\hline 
Basic reproduction number & ${\cal R}_{0}$  & Expected number of infections a typical infected individual can cause
in a fully susceptible population, ${\cal R}_{0}=Z_{x}T$.\tabularnewline
\hline 
\end{tabular}
\end{centering}
\end{table}

\begin{table}[ht]
\caption{Notation}
\begin{centering}
\begin{tabular}{|p{4cm}|p{1.5cm}|p{8cm}|}
\hline 
&  & \tabularnewline
Quantity  & Symbol  & Description\tabularnewline
&  & \tabularnewline
\hline 
Rate of new infections  & $\tilde{J}(t)$ \footnotemark[1]  & $\tilde{J}(t)\delta t$ gives the number of new infections between
times $t$ and $t+\delta t$.\tabularnewline
\hline
 & $\tilde{\Theta}(\tau,t)$ & Fraction of active S-I edges where disease is actually transmitted
exactly at time $t$.\tabularnewline
\hline 
\# of infectious individuals  & $\tilde{I}(t)$  & Total number of infectious individuals at time $t$, $\tilde{I}(t)=\int_{0}^{t}\tilde{J}(t-\tau)\tilde{\Psi}(\tau,t)d\tau$.\tabularnewline
\hline 
\# of removed individuals  & $\tilde{R}(t)$  & Total number of removed individuals at time $t$, $\tilde{R}(t)=\int_{0}^{t}\tilde{J}(t-\tau)[1-\tilde{\Psi}(\tau,t)]d\tau$.\tabularnewline
\hline 
 & $\tilde{R}^{r}(t)$  & Total number of removed individuals at time $t$ whose predecessor
is already removed.\tabularnewline
\hline 
 & $\tilde{R}^{i}(t)$  & Total number of removed individuals at time $t$ whose predecessor
is still infectious.\tabularnewline
\hline 
 & $\tilde{I}^{r}(t)$  & Total number of infectious individuals at time $t$ whose predecessor
is already removed.\tabularnewline
\hline 
 & $\tilde{I}^{i}(t)$  & Total number of infectious individuals at time $t$ whose predecessor
is still infectious.\tabularnewline
\hline 
 & $\tilde{{\cal Z}}_{x}^{r}(t)$ & Total number of excess links of removed individuals, calculated by
Eq.~\eqref{eq:z_tilde}.\tabularnewline
\hline 
Transmissibility of removed individuals & $T^{r}(t)$  & Gives the transmissibility of removed individuals at time $t$ and
it is calculated by Eqs.~\eqref{TT} and \eqref{Tr}.\tabularnewline
\hline 
 & $\chi(\tau)$  & $\chi(\tau)\delta\tau$ gives the expected number of new infections
produced by an infectious individual between ages of infection $\tau$
and $\tau+d\tau$. \tabularnewline
\hline 
Generation interval distribution & $\hat{\chi}(\tau)$  & $\hat{\chi}(\tau)\delta\tau$ gives the conditional probability that
given an infection, it occurred between ages of infection $\tau$
and $\tau+d\tau$. \tabularnewline
\hline 
\end{tabular}
\par\end{centering}
\end{table}
\end{document}